# A single-phase epitaxially grown ferroelectric perovskite nitride

**Authors:** Songhee Choi[1,†], Qiao Jin[1,†], Xian Zi[2,†], Dongke Rong[1,†], Jie Fang,[3] Jinfeng Zhang,[3] Qinghua Zhang[1], Wei Li[4], Shuai Xu[1], Shengru Chen[1,5], Haitao Hong[1,5], Cui Ting[1,5], Qianying Wang[1,5], Gang Tang[6], Chen Ge[1], Can Wang[1,5], Zhiguo Chen[1], Lin Gu[4], Qian Li[4], Lingfei Wang[3], Shanmin Wang[7,8,*], Jiawang Hong[2,*], Kuijuan Jin[1,5,*], and Er-Jia Guo[1,5,*]

**Affiliations**

[1] Beijing National Laboratory for Condensed Matter Physics and Institute of Physics, Chinese Academy of Sciences, Beijing 100190, China

[2] School of Aerospace Engineering, Beijing Institute of Technology, Beijing 100081, China

[3] Hefei National Research Center for Physical Sciences at Microscale, University of Science and Technology of China, Hefei 230026, China

[4] School of Materials Science and Engineering, Tsinghua University, Beijing 100084, China

[5] Department of Physics & Center of Materials Science and Optoelectronics Engineering, University of Chinese Academy of Sciences, Beijing 100049, China

[6] Beijing Institute of Technology, Zhuhai Beijing Institute of Technology (BIT), Zhuhai 519088, China

[7] Department of Physics, Southern University of Science and Technology, Shenzhen 518055, China

[8] Quantum Science Center of Guangdong-Hongkong-Macao Greater Bay Area, Shenzhen, Guangdong, 518045, China

†These authors contribute equally to the manuscript.
*Corresponding author Emails: kjjin@iphy.ac.cn, wangsm@sustech.edu.cn, hongjw@bit.edu.cn, and ejguo@iphy.ac.cn

**Abstract**

The integration of ferroelectrics with semiconductors is crucial for developing functional devices, such as field-effect transistors, tunnel junctions, and nonvolatile memories. However, the synthesis of high-quality single-crystalline ferroelectric nitride perovskites has been limited, hindering a comprehensive understanding of their switching dynamics. Here we report the synthesis and characterizations of epitaxial single-phase ferroelectric cerium tantalum nitride ($CeTaN_3$) on both oxides and semiconductors. The polar symmetry of $CeTaN_3$ was confirmed by observing the atomic displacement of central ions relative to the center of the $TaN_6$ octahedra, as well as through optical second harmonic generation. We observed switchable ferroelectric domains in $CeTaN_3$ films using piezoresponse force microscopy, complemented by the characterization of square-like polarization-electric field hysteresis loops. The remanent polarization of $CeTaN_3$ reaches approximately 20 μC/cm² at room temperature, consistent with theoretical calculations. This work establishes a vital link between ferroelectric nitride perovskites and their practical applications, paving the way for next-generation information and energy-storage devices with enhanced performance, scalability, and manufacturability.

**Teaser** Single-phase ferroelectric $CeTaN_3$ films were integrated on semiconductor substrates for next-generation transistor and sensors.

**Main text**
**Introduction**

Integration with silicon is pivotal for the practical implementation of ferroelectrics (*1-5*). This integration offers dual benefits: firstly, it allows for the amalgamation of polarized state switching with the well-understood electrical characteristics of silicon, including high mobility, tunable carrier density, and chemical stability (*6,7*). These hybrid structures unlock enhanced functionalities beyond what silicon alone can achieve. Secondly, this integration is compatible with CMOS technology, enabling the utilization of existing infrastructure, cost reduction, and acceleration of device and technology developments (*8,9*). Presently, the integration of ferroelectric functionalities in close proximity to silicon relies on perovskite and binary oxides (*10-13*). There is a compelling need to explore other inorganic oxygen-free ferroelectric materials that exhibit comparable ferroelectric properties.

In contrast to perovskite oxides, there have been relatively few reports on nitride perovskites. As shown in Fig.1A, $ThTaN_3$ was the first theoretically predicted and experimentally discovered nitride perovskites, dating back to 1995 (*14*). $ThTaN_3$ powder was synthesized by reacting oxide precursors under nitrogen-rich environment at high temperatures ranging from 1100 to 1500°C. Following this pioneering work, there was a hiatus in the exploration of nitride perovskites due to synthesis challenges and the lack of remarkable physical properties in the initial precursor. Recently, many theoretical groups have revisited nitride perovskites and predicted that several of these materials may exhibit excellent ferroelectric and piezoelectric properties (*15-18*). For instance, Sarmiento-Pérez *et al.* used high-throughput calculations to predict numerous thermodynamically stable and experimentally accessible nitride perovskites (*15*). Among these, lanthanum tungsten nitride ($LaWN_3$) was identified as a potential ferroelectric semiconductor with a distorted perovskite structure in its ground state. An experimental study by Talley *et al.* reported that $LaWN_3$ polycrystalline thin films exhibit a polar structure and piezoelectric response comparable to oxides (*19*). Shortly thereafter, a group of nitride perovskites ($CeMoN_3$, $CeWN_3$, and *etc.*) was theoretically predicted and experimentally achieved (*20,21*). They used magnetron sputtering and $N_2$ annealing process to fabricate phase-pure Ce-based nitride perovskites that exhibit interesting optoelectronic responses and low-temperature magnetic behaviors. These pioneer-work initiated research on nitride perovskites, making them a material system as important as perovskite oxides and other functional materials.

The field of functional nitride perovskites research remains largely uncharted, and the recent experimental breakthrough challenges the focus on superior piezoelectric properties solely within oxides, sparking increased interest in nitride materials with a perovskite structure. The simplistic framework of the perovskite crystal structure enables the possibility of epitaxial growth of nitride perovskites on conventional oxide substrates, with (pseudo-)cubic lattice constants ranging from 3.8 to 4.2 Å (*22*). An even more exciting aspect is that the experience in the oxide interface engineering can be readily applied to the research of nitride perovskites. Simultaneously, the industry has developed mature techniques for integrating perovskite or binary oxides with silicon substrates, offering an opportunity to integrate nitride perovskites as well. This integration could have a profound impact on both fundamental research and applied technologies in future.

While the initial demonstration of polar nitride perovskites showcases a piezoelectric character suitable for capacitors and actuators, its micrometer-thickness, and polycrystalline nature currently restrict broader applications in ferroelectric devices (*19, 23*). Ideally, the high-quality single-crystalline thin films are preferred. The influence of grain boundaries is largely removed, resulting in the reduced leakage current and fast-switching dynamics (*24, 25*). Still, there is limited attempts for exploring the single-crystalline nitride perovskites thin films due to the difficulty in the synthesis. To fabricate the high-quality and correct stoichiometric nitride perovskites with negligible defects remains highly challenging, not to mention the integration with wide range of oxide and semiconductor substrates.

In this study, we synthesized first-ever single-crystalline thin films of cerium tantalum nitride ($CeTaN_3$), a material previously predicted theoretically but never experimentally achieved. Multiple techniques and electrical measurements were employed to confirm the polar structural symmetry and ferroelectric nature of $CeTaN_3$. At room temperature, its remnant polarization was found to be comparable to that of hafnium zirconium oxide and barium titanate, in good agreement with our first-principles calculations. The successful integration of high-quality perovskite nitride thin films on both oxide and semiconductor substrates underscores the potential of the broader nitride perovskites family. With its superior functionalities, this family emerges as a promising counterpart to the well-studied oxide, halide, and chalcogenide perovskites ([26-28]).

**Results**

**Growth of high-quality single-phase $CeTaN_3$ thin films**

The main challenge in stabilizing nitride perovskites is compensating for the three nitrogen ions ($N^{3-}$) to achieve a collective anion valence of -9, which must match the collective cation valence of $A^{4+}$ and $B^{5+}$ (or $A^{3+}$ and $B^{6+}$) in the composition of $ABN_3$. Earlier work by Ha *et al.* ([20]) computationally screened all possible nitride perovskites through comprehensive *ab initio* calculations, theoretically predicting $CeTaN_3$ as a stable semiconductor. Thus, we started from performing the first-principles calculations based on density functional theory (DFT) to compute the density of states (DOS) (Fig. 1B) and the minimized energy of its polar structure (Fig. 1C). The detailed band structure is provided in fig. S1. The band structures indicate that $CeTaN_3$ is a semiconductor with indirect band gap of 1.06 eV. We observed that the top of the valence band mainly arises from the orbital hybridization of Ce *f* and N *p* orbitals, while the Ce *f* orbitals also contribute to the bottom of the conduction band. With suitable substrate selection or cation doping, the bandgap of $CeTaN_3$ can be tuned by altering its phase and symmetry ([20]). Its relatively small bandgap allows absorption in the infrared and visible spectral range, making it a promising candidate for optoelectronic applications. Using DFT calculations, we further explored the evolution of the band gap in $CeTaN_3$ as a function of biaxial strain (figs. S2 and S3). With increasing compressive strain, the band gap widens due to the upward shift of the conduction band. To predict the possible ferroelectric structures of $CeTaN_3$, we had performed calculations for all space groups that meet the requirements of point group 4*mm* (based on the follow-up experimental results), and only *P*4*mm* was found to be stable. We investigated the energy stability of $CeTaN_3$ films with both centrosymmetric non-polar structures (P4/*mmm* phase) and polar structures (P4*mm* phase). The polar and non-polar structures are depicted in Fig. 1C. The ferroelectricity of tetragonal $CeTaN_3$ was demonstrated along the ferroelectric switching path from the non-polar phase to the polar phase. We present the calculated double well energy profiles as a function of calculated polarization, with energy barriers of ~0.9 eV/f.u., yielding to a large remnant polarization of ~26.6 μC/cm². The calculation results indicate that the ferroelectricity in $CeTaN_3$ is driven by the orbital hybridization of A-site cations and *p* orbitals of anions. This result is consistent with ferroelectrics that have lone pair electrons, such as $PbTiO_3$ and $BiFeO_3$, although $CeTaN_3$ does not possess lone pair electrons. Similar hybridization is also observed in other ferroelectric nitride perovskites ([14-18]), which contrasts with ferroelectric perovskite oxides without lone pair electrons, where ferroelectricity originates from the hybridization of *d* orbitals of cations and *p* orbitals of oxygen ions. We calculated the formation energies, cleavage energies, and surface energies of bulk $CeTaN_3$ with different crystallographic orientations (Fig. 1D and fig. S4). The notably low formation energy and strong stability, attributed to the harmonious coexistence of $Ce^{4+}$ and $Ta^{5+}$ oxidation states, suggest the feasibility of synthesizing $CeTaN_3$ single crystals via physical vapor deposition.

We used a high-pressure and high-temperature synthesis approach to prepare a stoichiometric $CeTaN_3$ ceramic target (fig. S5). This method has been effectively adopted in our previous works for various high-quality binary transition metal nitrides (*23, 29-31*) and was described in a recent report on triclinic $LaReN_3$ (*32*). The $CeTaN_3$ target is composed of Ce, Ta, and N with a molar ratio of approximately 1:1:3, as determined by X-ray diffraction (XRD) and energy-dispersive X-ray (EDX) measurements. Initially, we fabricated $CeTaN_3$ thin films on (001)-oriented $SrTiO_3$ single crystal substrates using pulsed laser deposition (PLD). The relatively high vacuum environment together with home-built flowing nitrogen plasma (generation of highly active nitrogen atoms) has largely removed the impact of remaining oxygen atoms during thin-film deposition. The as-grown $CeTaN_3$ films were amorphous and lacked apparent diffraction peaks; however, they became crystallized after a rapid thermal process (RTP) (see Methods). Previously, the same procedure was employed to synthesize crystallized $CeMoN_3$, $CeWN_3$, and $LaWN_3$ films using magnetron sputtering (*19, 21*). This method was found to effectively eliminate secondary fluorite-type phase. In our work, the slow growth rate by nitrogen-plasma assisted PLD largely preserved the chemical composition of the ceramic target. Post-deposition annealing in flowing ammonia using RTP further facilitated the formation of single-crystalline $CeTaN_3$ thin films while ensuring a pure single-phase perovskite. Following the same procedure, we also grew single-crystalline $CeTaN_3$ thin films on silicon substrates with 10-unit-cell-thick $SrTiO_3$ buffer layers. During the deposition, a 2–3 nm thick amorphous $SiO_2$ interlayer was naturally formed between the $SrTiO_3$ and silicon substrates (see schematic in Fig. 2A). Infrared absorption spectra of both amorphous and crystallized $CeTaN_3$ films, compared to bare $SrTiO_3$-buffered silicon substrates, were measured (fig. S6). The infrared absorption peak shifted from 300.9 to 263.3 $cm^{-1}$, as indicated by colored arrows, and the peak intensities at 536.7 and 611.4 $cm^{-1}$ slightly increased after crystallization. These peak shifts after crystallization suggest reduced phonon scattering in the crystallized $CeTaN_3$ films.

**Structural characterizations of high-quality single-crystalline $CeTaN_3$ thin films**

The structure of $CeTaN_3$ thin films on (001)-oriented $SrTiO_3$ and $SrTiO_3$/Si substrates were examined by XRD and scanning transmission electron microscopy (STEM) (figs. S7 to S10). Fig. 2B show the XRD $\theta$-$2\theta$ scans of $CeTaN_3$ thin films. The $CeTaN_3$ thin films are (00*l*)-oriented without other impurity peaks. The sharp diffraction peaks with clear Laue oscillations (inset of Fig. 2B) demonstrate the high crystallinity of $CeTaN_3$. The thickness of $CeTaN_3$ thin films is approximately 80 nm, determined from X-ray reflectivity measurements. The phi-scans around the 022 reflection of the $CeTaN_3$ films and 011 reflection of the Si substrates were measured (fig. S11). The results validate that the $CeTaN_3$ thin film was epitaxially grown on Si substrates with a 45° rotation along the surface normal, consistent with previous reports of perovskite oxides directly grown on silicon substrates (*33*). Analysis of the reciprocal space mapping (RSM) in Fig. 2C confirms that the $CeTaN_3$ films were epitaxially grown on silicon substrates. The $CeTaN_3$ film peak exhibits a slight shift to a lower $q_x$ position compared to that of the silicon substrates, indicating that the $CeTaN_3$ films are not fully strained and have slightly relaxed lattice constants. The RSM around the 013 reflection shows a single peak, indicating that the $CeTaN_3$ films show a uniform single-domain. We further determined that the $CeTaN_3$ films have a tetragonal-like structure with lattice constants $a = b = 4.03$ Å and $c = 4.09$ Å (fig. S12). Note that the tetragonality of $CeTaN_3$ films on Si substrates is slightly increased compared to the $CeTaN_3$ films grown directly on SrTiO$_3$ substrates. The chemical composition of CeTaN$_3$ thin films was determined by X-ray photoelectron spectroscopy (XPS) measurements (fig. S13). The single oxidation states of Ta$^{5+}$ and N$^{3-}$ ions were confirmed, while the main peaks are attributed to Ce$^{4+}$ with a minor presence of Ce$^{3+}$, likely due to inevitable small amount of nitrogen vacancies or dislocations present in CeTaN$_3$ surface layers. The occupied electronic state around Fermi level for CeTaN$_3$ on SrTiO$_3$/Si were taken from XPS valence band (VB) spectra (fig. S14). The DOS across Fermi level matches well with the insulating behavior of CeTaN$_3$ films. The broad peak around 4–8 eV is mostly formed by N 2$p$ bands and hybridized bands between cations and nitrogen ions. The obtained band gap of CeTaN$_3$ thin films is ~1.27 eV. In addition, we performed optical absorption spectroscopy measurements (fig. S14). The results show that the absorption coefficient (α) is proportional to ($\hbar\omega - E_g$)$^2$, indicating that CeTaN$_3$ is an indirect bandgap semiconductor. The extracted optical bandgap is approximately 1.16 eV, consistent with our former theoretical calculations (Fig. 1B). To assess the chemical uniformity of CeTaN$_3$ thin films, we performed time-of-flight secondary ion mass spectrometry (ToF-SIMS) measurements. The results confirm the uniformity of all deposited layers and provide insight into the chemical distribution from top to bottom (Fig. S15).

The atomic-scale structural coherency was examined by high angle annular dark field (HADDF) and annular bright field (ABF) imaging taken in scanning transmission electron microscopy (STEM) mode. High-resolution cross-sectional STEM image of CeTaN$_3$/SrTiO$_3$/$a$-SiO$_2$/Si was viewed along silicon substrate's [110] and [010] orientations (Fig. 2D and fig. S16). These images demonstrate the epitaxial growth of CeTaN$_3$ thin films with chemically sharp and coherent interfaces (indicated by yellow dashed lines). The selected area electron diffraction (SAED) pattern was acquired from the interface region, as shown in Fig. 2E. There are two series of sharp and periodic diffraction spots from SAED patterns. In the cross-section view, the blue circled peaks (CeTaN$_3$) show square while the yellow circled peaks (Si) represent rectangle with long transverse than longitudinal in the reciprocal space. We found that the out-of-plane CeTaN$_3$ [010] ∥ Si [010] alignment and in-plane CeTaN$_3$ [100] ∥ Si [100] alignment. To characterize the chemical composition of CeTaN$_3$, the STEM electron energy lose spectra (EELS) at a representative region in CeTaN$_3$ were measured. Figs. 2G to 2I show the atomical positions of each Ce, Ta, and N atoms with chemical compositions. From STEM-EELS results, we could clearly identify the perovskite structure of CeTaN$_3$ and uniformed chemical distribution. Fig. 2J shows the close-to-background signal from O $K$-edges, suggesting the extremely low concentration of oxygen in CeTaN$_3$ films. The crystal-homogeneity of CeTaN$_3$ was confirmed in expanded atomic-resolution STEM images (Fig. 2F and fig. S17). The local polarization within the CeTaN$_3$ lattice can be quantitatively determined on the basis of the displacement of TaN$_6$ octahedral relative to the center of the TaN$_6$ octahedral (*34*). The CeTaN$_3$ films show typical polar domains with displacement vectors pointing towards the same direction. The typical polar domain structures with diameters of tens of μm and domain volume fraction of approximately 70%. The dashed regions indicate the polar domains within the CeTaN$_3$ films. The polarization of domains points towards upwards with both in-plane and out-of-plane vectors for the intrinsic states. Please note that the films are not completely polarized by electric field, meaning that the observed polar domains are the imprints within the films. Therefore, the polarization is supposed to be relatively smaller than the polarized states. fig. S18 shows the ABF image of CeTaN$_3$ films acquired in the same region. The Ta atoms were shifted from the center of TaN$_6$ octahedra, resulting in polar symmetry breaking. We extracted the averaged displacement of Ta atoms is 12.1 ± 1.3 pm from the center of the TaN$_6$ octahedral along the [001] orientation from the cubic center. Thus, from the atomic displacement, we calculated the ferroelectric polarization of CeTaN$_3$ is approximately 20-25 μC/cm$^2$ using the Born effective charges.

**Confirmation of polar structure and switchable piezoresponse in CeTaN$_3$**

We conducted second harmonic generation (SHG) measurements on CeTaN$_3$ thin films grown on SrTiO$_3$/Si substrates to shed light on the structural origin of ferroelectrics (see setup in Fig. 3A). This technique is a sensitive probe for the structure symmetry and polar nature. Fig. 3B shows the experimental data and theoretical fits to the SHG intensity $I_{P-out}^{2\omega}$ and $I_{S-out}^{2\omega}$. We verified that CeTaN$_3$ films have *P4mm* space group, exhibiting stable polar symmetry character. In contrast, the amorphous CeTaN$_3$ films do not show apparent SHG signals. The uniformity of polar domains was examined using lateral SHG mapping (Fig. 3C). The brightness of SHG image represents higher SHG intensity, *i.e.* larger polarization. The CeTaN$_3$ films exhibit bright SHG signal spots randomly distributed with clear boundaries. Apparently, the ferroelectric domains are unevenly distributed over a 50 × 50 μm$^2$ region, in agreement with previous STEM observations. The size of polar domains is approximately in the micrometer scale.

The macroscopic analysis for polarization symmetry breaking and electrical piezoresponse of CeTaN$_3$ was conducted with piezoresponse force microscopy (PFM). Figs. 3D and 3E display the local piezoelectric hysteresis loops as a function of tip voltage in both phase and amplitude signals, respectively. A 15-nm-thick CeTaN$_3$ film exhibits polarization switching at approximately ±5 V, while the switching voltages decrease for an 80-nm-thick CeTaN$_3$ film. Notably, we observed asymmetric switching voltages in the thicker CeTaN$_3$ films (+2 V for downward switching and –4.4 V for upward switching). The increased coercive field in thinner ferroelectric films arises from the increase nucleation energy barrier and enhanced depolarization field. Additionally, the piezoresponse amplitude increases with film thickness. Figs. 3F and 3G show PFM phase images from the 15-nm-thick and 80-nm-thick CeTaN$_3$ films, respectively, with full sets of PFM amplitude and phase images provided in figs. S19 and S20. To conduct the measurements, the sample was first poled with a fixed tip voltage over a large square area of 4×4 μm². Then, the characters "CeTaN$_3$" were written using the same tip voltage but with opposite polarity. The switching voltages progressively increased from left to right. For the 15-nm-thick CeTaN$_3$ film, no stable domain could be written at a tip voltage below 8 V, possibly due to unswitchable domains at low fields or domain relaxation between writing and reading. Clear written domains were observed after increasing the tip voltage to 12 V. The same procedure was repeated on the 80-nm-thick CeTaN$_3$ film, and due to its lower switching field, stable written domains appeared even at ±8 V. It is noteworthy that the domain size, i.e., the width of the written characters, increased with higher switching voltages—a common phenomenon in ferroelectric oxides, where domain walls expand laterally faster at higher tip voltages and fixed duration time (scanning speed). All measurements were repeated with reversibly switched tip voltages at multiple positions on the film surface. Remarkably, the written domain patterns remained clearly identifiable after several weeks, confirming the non-volatile nature of the switchable domains.

**Universal synthesis procedure for single-crystalline CeTaN$_3$ films on semiconductors**

To demonstrate the universal synthesis procedure, we fabricated CeTaN$_3$ films directly on Si substrates, as well as on other third-generation semiconductor substrates, including SiC and GaN (Figs. 4A to 4C). As shown in Fig. 1D, the formation energy of (011)-oriented CeTaN$_3$ is lower than that of (001)-oriented CeTaN$_3$. Therefore, in the absence of substrate regulation, the CeTaN$_3$ films preferentially crystallize along the (011) orientation. Fig. 4E displays the XRD *θ-2θ* scans of CeTaN$_3$ thin films on different substrates. Only (0*ll*) peaks from CeTaN$_3$ and peaks from the substrates (indicated by '*') were observed. A direct comparison with CeTaN$_3$ single-crystalline thin films grown on (011)-oriented SrTiO$_3$ substrates confirms that the (011)-oriented CeTaN$_3$ films were epitaxially grown on all substrates. Microstructural analysis further confirms the epitaxial growth of CeTaN$_3$ thin films on various substrates (figs. S21 and S22). Due to the base pressure of the high-temperature PLD growth chamber reaching only ~1×10$^{-8}$ Torr, oxidation of the Si substrate was unavoidable, leading to the formation of an ultrathin SiO$_2$ layer, approximately 1 nm thick, between the CeTaN$_3$ film and the Si substrate (fig. S21). Additionally, detailed EDX mapping of a CeTaN$_3$ film grown on Si (fig. S23) shows uniform distribution of Ce, Ta, and N signals within the films, with no apparent oxygen content detected. Chemical intermixing at the CeTaN$_3$/Si interface was negligible.

Fig. 4F presents a representative STEM-ABF image of (011)-oriented CeTaN$_3$ films, where the light dark spots correspond to N atoms. The in-phase rotation along the [001] orientation (*b*-axis) allows for direct visualization and quantification of TaN$_6$ octahedral tilts in the *ac*-plane. Owing to the high crystallinity of CeTaN$_3$ thin films, this work, to our knowledge, is the first to reveal nitrogen octahedral tilts in nitride perovskites. The Ta-N-Ta bond angle ($\beta_{\text{Ta-N-Ta}}$) reaches approximately 158° and the TaN$_6$ octahedra tilt is approximately 11°. The octahedral tilt remains nearly constant within the observed region. This small tilt angle may result from substrate-induced compressive strain. To examine the effects of misfit strain on the tilt angle, we fabricated CeTaN$_3$ thin films on various oxide substrates with progressively decreasing in-plane lattice constants (figs. S24 and S25). Our results show that the $\beta_{\text{Ta-N-Ta}}$ angle decreases with increasing compressive strain, consistent with previous studies on oxide heterostructures. The EELS mapping results for a CeTaN$_3$ thin film on DyScO$_3$ (fig. S26) reveal exceptionally high chemical homogeneity, with minimal intermixing at the heterointerface, confined to a single unit cell. The pristine interface between the nitride perovskite and oxide is attributed to their chemical stability, structural compatibility, and well-defined interface termination, exemplified by the TaN$_2$–DyO configuration at the CeTaN$_3$/DyScO$_3$ interface.

The piezoresponse of CeTaN$_3$ thin films on Si was confirmed by poling quadratic areas using oppositely written voltages (fig. S27). Macroscopic electric measurements were conducted at room temperature. Fig. 4G shows the electrical hysteresis loops (*P-E* and *J-E* curves) of a Pt/CeTaN$_3$/*p*-Si capacitor. The Pt/Ti top electrode area was reduced to 5 μm in diameter to minimize leakage current to the order of sub-μA. The remanent polarization ($P_r$) and the coercive field ($E_C$) of CeTaN$_3$ films are approximately 17.4 μC/cm$^2$ and 484 kV/cm, respectively. To confirm the intrinsic ferroelectric nature of CeTaN$_3$ thin films, we performed positive-up-negative-down (PUND) measurements (fig. S28) and calculated the switched polarization under different writing pulses (fig. S29). The switching current varies with the applied electric field and writing time. The characteristic switching time was determined to be ~1×10$^{-4}$ s at 667 kV/cm. The higher fields accelerating ferroelectric domain switching, where the smaller fields decrease the ferroelectric switching and lead to incomplete domain switching when fields below $E_C$ (fig. S30). Please note that both $P_r$ and switching speed are comparable to those of typical ferroelectric oxides (such as Hf$_{1-x}$Zr$_x$O$_2$ and BaTiO$_3$) (*37,38*). The $E_C$ of CeTaN$_3$ films is also comparable to those of the conventional ferroelectric perovskite oxides (*39-41*), atomically thin binary oxides (*42*), two dimensional materials (*43*), and elementary substances (*44*). Further investigations into the scaling of ferroelectric parameters with respect to thickness/orientation and switching dynamics in both microscopic scales are essential for advancing the potential of CeTaN$_3$ in ferroelectric applications with semiconductor industry in future.

**Discussion and Conclusions**

In conclusion, we have successfully achieved the first-ever single-crystalline ferroelectric CeTaN$_3$ thin films on silicon substrates, demonstrating the potential for realizing comparable ferroelectric polarization in nitride perovskites. The spontaneous remanent polarization and intrinsic ferroelectric atomic displacement within the TaN$_6$ octahedra reveal a strong correlation between the electronic structure and inversion symmetry. The experimentally observed ferroelectric order parameters align excellently with theoretical calculations. The clear observation of nitrogen octahedral tilting in nitride perovskites, reported here for the first time, highlights the sensitivity of nitrogen octahedra to epitaxial strain. Fine-tuning of octahedral parameters—such as rotation, bond angles, and bond lengths—could enable quantum confinements of their emergent physical properties. Additionally, the similarity in octahedral engineering between oxide and nitride perovskites suggests that combining these materials could open a largely unexplored research avenue at the oxide/nitride interfaces. Our work also emphasizes the potential for integrating single-crystalline CeTaN$_3$ thin films with second- and third-generation semiconductor substrates, as well as perovskite oxide substrates. This integration enhances the advantages of this ferroelectricity in nitride perovskites, not only in fundamental research but also for next-generation AI-driven computing and memory devices. Furthermore, we noticed that a recent publication reported the Ruddlesden–Popper nitrides, specifically Ce$_2$TaN$_4$, were synthesized under high pressure (*45*). This compound exhibited long-range order in charge distribution, off-center displacement, octahedral tilting, and spin states, suggesting that this material system holds potential as a functional nitride. Consequently, the exploration of not only single-crystalline perovskite nitrides but also Ruddlesden–Popper nitrides—and indeed a broader array of nitride materials—remains an exciting frontier for future research.

**Materials and Methods**

**CeTaN$_3$ target preparation and thin film growth**

Polycrystalline CeTaN$_3$ serves as ablation target with 25 mm in diameter and 5 mm in thickness. The target was sintered at 3 GPa and 1000 ºC for 60 min, using the single-phase perovskite nitride powders as starting materials. The nitride powders were synthesized though a recently formulated high-pressure reaction route. High-pressure experiments were carried out in a DS 6 × 10 MN cubic press installed at the High-Pressure Lab of South University of Science and Technology (SUSTech). The CeTaN$_3$ target was used to fabricate the thin films using pulsed laser deposition system with a XeCl excimer laser of 308 nm wavelength and 5 Hz repetition. The laser spot size is approximately 5.5 mm$^2$ and the energy density maintains ~1 J/cm$^2$ during the deposition. We employed highly active nitrogen atom generator-nitrogen plasma to compensate the nitrogen vacancies during the deposition. The (001)- and (011)-oriented SrTiO$_3$, (001)-oriented LaAlO$_3$, (110)-oriented DyScO$_3$, and *n*-type Si, SiC, GaN/Al$_2$O$_3$ single crystals were used as substrates. A 5-nm-thick SrTiO$_3$ buffer layer was prepared on Si substrates using two step methods which described in our previous works (46,47,48). The CeTaN$_3$ thin films were prepared in two steps. Firstly, the as-grown stoichiometric films were deposited at the substate temperature of 600 °C and the base pressure of 1 × 10$^{-8}$ Torr. The thickness of CeTaN$_3$ thin films was controlled by counting the number of laser pulses. The second step was to anneal the films using rapid thermal process (RTP), which is crucial for crystalizing the CeTaN$_3$ thin films. The annealling process was performed by heating up the films to 800 °C in 10 mins under pure anomia atmosphere. After 30 mins annealling, the films cooled down to room temperature slowling.

**Structure characterizations**

The thin film quality was examined by high-resolution four-circle X-ray diffractormetor (XRD, Malvern Panalytical, X'Pert3 MRD) and synchrotron-based XRD at the beamline 1W1A of the Beijing Synchrotron Radiation Facility (BSRF). The θ-2θ line scans, rocking curve scans, φ scans, and off-specular RSM were conducted to check the epitaxial growth of CeTaN$_3$ thin films. The atomic structure of CeTaN$_3$ thin films were characterized using an ARM-200CF scanning transmission electron microscope (STEM) which was operated at 200 keV and was equipped with double spherical aberration (Cs) correctors. The STEM spicemen were prepared by standard mechanical thinning followed by ion-milling. All TEM images were processed and analyzed using Gatan DigitalMicrograph.

**Time-of-Flight (ToF) Secondary Ion Mass Spectrometry (ToF-SIMS) measurements**

ToF-SIMS measurements were carried out using a ToF-SIMS (ION-TOF GmbH, M6 instrument,Münster, Germany). The mass spectrometer was equipped with a reflection type ToF analyzer. A dual-beam depth profiling strategy was employed, in which a 1 keV Ar ion Gun (300 μm × 300 μm scanning area) was used for sputtering and a 30 keV Bi$_3^+$ beam (~0.43 pA, 100 μm × 100 μm scanning area within the center of the Cs+ crater) was used for negative spectra data collection with high mass resolution mode. Additionally, a flood gun (~1 μA) was used for charge compensation. The CeTaN$_3$film/SrTiO$_3$/Si substrate interfaces were determined via the secondary ion signals of SrO$^+$ and Sr$^{4+}$.

**Optical reflectance measurements**

The optical reflectance measurements in the energy range from 10 to 1000 meV were performed on a Bruker Vertex 80v Fourier transform spectrometer at room temperature. The transmitance spectra of amorphous and crystallized CeTaN$_3$ thin films on SrTiO$_3$-capped Si substrates were measured, in comparison to that of a bare SrTiO$_3$-capped Si substrates. The optical transmitance data were highly reproducble. Moreover, the optical constants of the CeTaN$_3$ thin films were obtained using a ellipsometer (J.A. Woollam RC2 spectroscopy), which are consistant with the results extracted from the measured reflectance spectra in the same energy range.

**Optical second harmonic generation (SHG)**

Optical SHG measurements were performed on two independent home-developed systems. For the point SHG measurements, we performed at the Institute of Physics, CAS. A Ti: Sapphire femtosecond laser (Tsunami 3941-X1BB, Spectra-Physics) with a pulse duration of 120 fs, a center wavelength of 800 nm, and an average power of 50-200 mW was employed as light source. The 1ω laser beam were incident on the CeTaN$_3$ thin films with an angle of 45°, while the 2ω signals in the reflected laser beam were band-pass filtered and recorded with a photodetector. For the SHG mapping measurements, the experiment was conducted at Tsinghua University using a laser-scanning microscope system. We employed a normal incidence geometry. The 1ω laser beam was focused by an objective lens and scanned with a pair of galvanometers. The backscattered 2ω light signals were collected from reflection beam using a photomultiplier tube. In both cases, the polarization of incident beam was controlled with a zero-order half-wave plate, the polarization of the SHG light was analyzed using a Glan-Taylor prism. The intensities of *p*- and *s*-polarized SHG light were plotted as a function of beam polarization, *i. e.* the angle with respect to beam. All SHG measurements were performed at room temperature.

**Ferroelectric characterizations**

Piezoresponse force microscopy (PFM) were conducted using Asylum, MFP-3D at room temperature. A Si cantilever with conductive Pt/Ir coating layer of spring constant of 0.2 N/m was used. Both amplitude and phase signals were recorderd. The DC bias voltages were applied from 4 to 12 V for switching ferroelectric domains. Ferroelectric polarization measurements have been conducted with an Aixacct TF 2000 Analyzer on CeTaN$_3$ thin film capacitor structures with Pt top electrodes with an area as small as 25 μm$^2$ area in order to largely reduce the leakage current.

**X-ray photoelectron spectroscopy (XPS)**

Room-temperature X-ray photoelectron spectroscopy (XPS) measurements were performed at the Institute of Physics, CAS. All XPS measurements were measured with X-rays at a normal angle to the sample surface. The samples were chemically cleaned to remove the surface contaminations. Spectra were measured using an electron flood gun to compensate the positive photoemission charge because $CeTaN_3$ thin films were highly insulating. A small polycrystalline Au foil was affixed to the corner of each film surface using Cu tape in order to connect the sample to the ground. For valence band spectra, the Au $4f_{7/2}$ peak was used to calibrate the binding-energy scale. The distribution of element valence states was obtained by fitting N $2p$ XPS using Casa XPS software.

**Density-functional theory (DFT) calculations**

The calculations were performed in the framework of density functional theory (DFT) as implemented in the Vienna ab initio simulation package code (*49, 50*). The electron exchange-correlation effects were treated with the generalized gradient approximation (GGA) with Perdew-Burke-Ernzerhof solid functional (PBEsol) (*51, 52*). The Heyd-Scuseria-Ernzerhof (HSE06) functional was applied for the calculation of the density of states (DOS).(*51*) Energy convergence can be further confirmed by using a cutoff energy of 500 eV and a k-point grid with a reciprocal space resolution of 0.2 Å$^{-1}$. The in-plane lattice constant was fixed to the experimental values during relaxation, while the out-of-plane lattice constant and atomic position were fully relaxed. The PSEUDO program from the Bilbao Crystallographic Server (BCS) was employed to identify the centrosymmetric structure (*53-55*), while the VTST code was employed to obtain the insert structures along ferroelectric switching path from non-polar phase to polar phase (*56, 57*).

**References**
1. R. A. McKee, F. J. Walker, M. F. Chisholm, Crystalline oxides on silicon: the first five monolayers. *Phys. Rev. Lett.* **81,** 3014-3017 (1998).
2. A. Pasquarello, M. S. Hybertsen, R. Car, Interface structure between silicon and its oxide by first-principle molecular dynamics. *Nature* **396,** 58-60 (1998).
3. M. P. Warusawithana, C. Cen, C. R. Sleasman, J. C. Woicik, Y. Li, L. F. Kourkoutis, J. A. Klug, H. Li, P. Ryan, L.-P. Wang, M. Bedzyk, D. A. Muller, L.-Q. Chen, J. Levy, D. G. Schlom, A ferroelectric oxide made directly on silicon. *Science* **324,** 367–370 (2009).
4. S. H. Baek, J. Park, D. M. Kim, V. A. Aksyuk, R. R. Das, S. D. Bu, D. A. Felker, J. Lettieri, V. Vaithyanathan, S. S. N. Bharadwaja, N. Bassiri-Gharb, Y. B. Chen, H. P. Sun, C. M. Folkman, H. W. Jang, D. J. Kreft, S. K. Streiffer, R. Ramesh, X. Q. Pan, S. Trolier-McKinstry, D. G. Schlom, M. S. Rzchowski, R. H. Blick, C. B Eom, Giant piezoelectricity on Si for hyperactive MEMS. *Science* **334,** 958-961 (2011).
5. Lee, S. K., Choi, B. H. & Hesse, D. Epitaxial growth of multiferroic $BiFeO_3$ thin films with (101) and (111) orientations on (100) Si substrates. Epitaxial growth of multiferroic $BiFeO_3$ thin films with (101) and (111) orientations on (100) Si substrates. *Appl. Phys. Lett.* **102,** 242906 (2013).
6. M. Suzuki, Review on future ferroelectric nonvolatile memory: FeRAM. *J. Ceram. Soc. Jpn.* **103,** 1099-1111 (1995).
7. P. Vettiger, G. Binnig, The nanoderive project. *Sci. Am.* **288,** 46 (2003).
8. A. L. Esaki, R. B. Laibowitz, P. J. Stiles, Polar switch. *IBM Tech. Discl. Bull* **13,** 2161 (1971).
9. Y.-R. Wu, J. Singh, Polar heterostructure for multifunction devices: theoretical studies. *IEEE Trans. Electron Devices* **52,** 284-293 (2005).
10. S. S. Cheema, D. Kwon, N. Shanker, R. dos Reis, S.-L. Hsu, J. Xiao, H. Zhang, R. Wagner, A. Datar, M. R. McCarter, C. R. Serrao, A. K. Yadav, G. Karbasian, C.-H. Hsu, A. J. Tan, L.-C. Wang, V.

Thakare, X. Zhang, A. Mehta, E. Karapetrova, R. V Chopdekar, P. Shafer, E. Arenholz, C. Hu, R. Proksch, R. Ramesh, J. Ciston, S. Salahuddin, Emergence of room-temperature ferroelectricity at reduced dimensions. *Nature* **580,** 478-482 (2020).

11. S. S. Cheema, N. Shanker, L.-C. Wang, C.-H. Hsu, S.-L. Hsu, Y.-H. Liao, M. S. Jose, J. Gomez, W. Chakraborty, W. Li, J.-H. Bae, S. K. Volkman, D. Kwon, Y. Rho, G. Pinelli, R. Rastogi, D. Pipitone, C. Stull, M. Cook, B. Tyrrell, V. A. Stoica, Z. Zhang, J. W. Freeland, C. J. Tassone, A. Mehta, G. Saheli, D. Thompson, D. I. Suh, W.-T. Koo, K.-J. Nam, D. J. Jung, W.-B. Song, C.-H. Lin, S. Nam, J. Heo, N. Parihar, C. P. Grigoropoulos, P. Shafer, P. Fay, R. Ramesh, S. Mahapatra, J. Ciston, S. Datta, M. Mohamed, C. Hu, S. Salahuddin, Enhanced ferroelectricity in ultrathin films grown directly on silicon. *Nature* **604,** 65-71 (2022).

12. S. S. Cheema, N. Shanker, S.-L. Hsu, Y. Rho, C.-H. Hsu, V. A. Stoica, Z. Zhang, J. W. Freeland, P. Shafer, C. P. Grigoropoulos, J. Ciston, S. Salahuddin, Emergent ferroelectricity in subnanometer binary oxide films on silicon. *Science* **376,** 648-652 (2022).

13. K. J. Hubbard, D. G. Schlom, Thermodynamic stability of binary oxides in contact with silicon. *J. Mater. Res.* **11,** 2757-2776 (1996).

14. N. E. Brese, F. J. DiSalvo, Synthesis of the first thorium-containing nitride perovskite, $TaThN_3$. *J. Solid State Chem.* **120,** 378-380 (1995).

15. R. Sarmiento-Pérez, T. F. T. Cerqueira, S. Körbel, S. Botti, M. A. L. Marques, Prediction of stable nitride perovskites. *Chem. Mater.* **27,** 5957-5963 (2015).

16. Y.-W. Fang, C. A. J. Fisher, A. Kuwabara, X.-W. Shen, T. Ogawa, H. Moriwake, R. Huang, C.-G. Duan, Lattice dynamics and ferroelectric properties of the nitride perovskite $LaWN_3$. *Phys. Rev. B* **95,** 014111 (2017).

17. C. Gui, J. Chen, S. Dong, Multiferroic nitride perovskites with giant polarizations and large magnetic moments. *Phys. Rev. B* **106,** 184418 (2022).

18. X. Zi, Z. Deng, L. Rao, Y. Li, G. Tang, J. Hong, J. First-principles study of ferroelectric, dielectric, and piezoelectric properties in the nitride perovskites $CeBN_3$ (B = Nb, Ta). *Phys. Rev. B* **109,** 115125 (2024).

19. K. R. Talley, C. L. Perkins, D. R. Diercks, G. L. Brennecka, A. Zakutayev, A. Synthesis of $LaWN_3$ nitride perovskite with polar symmetry. *Science* **374,** 1488 (2021).

20. V.-A. Ha, H. Lee, F. Giustino, $CeTaN_3$ and $CeNbN_3$: prospective nitride perovskites with optimal photovoltaic band gaps. *Chem. Mater.* **34,** 2107-2122 (2022).

21. R. Sherbondy, R. W. Smaha, C. J. Bartel, M. E. Holtz, K. R. Talley, B. Levy-Wendt, C. L. Perkins, S. Eley, A. Zakutayev, G. L. Brennecka, High-throughput selection and experimental realization of two Ce-based nitride perovskites: $CeMoN_3$ and $CeWN_3$. *Chem. Mater.* **34,** 6883-6893 (2022).

22. A. Biswas, C.-H. Yang, R. Ramesh, Y. H. Jeong, Atomically flat single terminated oxide substrate surfaces. *Prog. Surf. Sci.* **92,** 117-141 (2017).

23. X. Zhou, W. Xu, Z. Gui, C. Gu, J. Chen, J. Xie, X. Yao, J. Dai, J. Zhu, L. Wu, E.-j. Guo, X. Yu, L. Fang, Y. Zhao, L. Huang, S. Wang, Polar nitride perovskite $LaWN_{3-\delta}$ with orthorhombic structure. *Adv. Sci.* **10,** 2205479 (2023).

24. P. Gao, C. T. Nelson, J. R. Jokisaari, S. H. Baek, C. W. Bark, Y. Zhang, E. Wang, D. G. Schlom, C. B. Eom, X. Pan, Revealing the role of defects in ferroelectric switching with atomic resolution. *Nat. Commun.* **2,** 591 (2011).

25. E.-J. Guo, R. Roth, A. Herklotz, D. Hesse, K. Dörr, K. Ferroelectric 180° domain wall motion controlled by biaxial strain. *Adv. Mater.* **27,** 1615-1618 (2015).

26. S. Geng, Z. Xiao, Can nitride perovskites provide the same superior optoelectronic properties as lead halide perovskites? *ACS Energy Lett.* **8,** 2051-2057 (2023).

27. E. Aydin, T. G. Allen, M. D. Bastiani, A. Razzaq, L. Xu, E. Ugur, J. Liu, S. D. Wolf, Pathways toward commercial perovskite/silicon tandem photovoltaics. *Science* **383,** 6679 (2024).

28. B. F. Grosso, D. W. Davies, B. Zhu, A. Walsh, D. O. Scanlon, Accessible chemical space for metal nitride perovskites. *Chem. Sci.* **14,** 9175-9185 (2023).

29. Q. Jin, H. Cheng, Z. Wang, Q. Zhang, S. Lin, M. A. Roldan, J. Zhao, J.-O. Wang, S. Chen, M. He, C. Ge, C. Wang, H.-B. Lu, H. Guo, L. Gu, X. Tong, T. Zhu, S. Wang, H. Yang, K.-j. Jin, E.-J. Guo, Strain-mediated high conductivity in ultrathin antiferromagnetic metallic nitrides. *Adv. Mater.* **33,** 2005920 (2021).

30. Q. Jin, Z. Wang, Q. Zhang, Y. Yu, S. Lin, S. Chen, M. Qi, H. Bai, A. Huon, Q. Li, L. Wang, X. Yin, C. S. Tang, A. T. S. Wee, F. Meng, J. Zhao, J.-o. Wang, H. Guo, C. Ge, C. Wang, W. Yan, T. Zhu, L. Gu, S. A. Chambers, S. Das, T. Charlton, M. R. Fitzsimmons, G.-Q. Liu, S. Wang, K.-j. Jin, H. Yang, E.-J. Guo, Room-temperature ferromagnetism at an oxide-nitride interface. *Phys. Rev. Lett.* **128,** 017202 (2022).

31. Q. Jin, Q. Zhang, H. Bai, M. Yang, Y. Ga, S. Chen, H. Hong, T. Cui, D. Rong, T. Lin, J.-O. Wang, C. Ge, C. Wang, Y. Cao, L. Gu, G. Song, S. Wang, K. Jiang, Z.-G. Cheng, T. Zhu, H. Yang, K.-j. Jin, E.-J. Guo, Syntropic spin alignment at the interface between ferromagnetic and superconducting nitrides. *Nat. Sci. Rev.* **11,** nwae107 (2024).

32. S. D. Kloß, M. L. Weidemann, J. P. Attfield, Preparation of bulk-phase nitride perovskite LaReN$_3$ and topotactic reduction to LaNiO$_2$-type LaReN$_2$. *Angew. Chem., Int. Ed.* **60,** 22260-22264 (2021).

33. S.-H. Baek, Eom C.-B. Epitaxial integration of perovskite-based multifunctional oxides on silicon. *Acta Mater.* **61,** 2734-2750 (2013).

34. A. K. Yadav, C. T. Nelson, S. L. Hsu, Z. Hong, J. D. Clarkson, C. M. Schlepütz, A. R. Damodaran1, P. Shafer, E. Arenholz, L. R. Dedon, D. Chen, A. Vishwanath, A. M. Minor, L. Q. Chen, J. F. Scott, L. W. Martin, R. Ramesh, Observation of polar vortices in oxide superlattices. *Nature* **530,** 198-201 (2016).

35. J. Wang, J. B. Neaton, H. Zheng, V. Nagarajan, S. B. Ogale, B. Liu, D. Viehland, V. Vaithyanathan, D. G. Schlom, U. V. Waghmare, N. A. Spaldin, K. M. Rabe, M. Wuttig, R. Ramesh, Epitaxial BiFeO$_3$ multiferroic thin film heterostructures. *Science* **299,** 1719-1722 (2003).

36. B. I. Vrejoiu, G. L. Rhun, L. Pintilie, D. Hesse, M. Alexe, U. Gösele, Intrinsic ferroelectric properteis of strained tetragonal PbZr$_{0.2}$Ti$_{0.8}$O$_3$ obtained on layer-by-layer grown, defect-free single crystalline films. *Adv. Mater.* **18,** 1657-1661 (2006).

37. J. F. Scott, Applications of modern ferroelectrics. *Science* **315,** 954-959 (2007).

38. K. J. Choi, M. Biegalski, Y. L. Li, A. Sharan, J. Schubert, R. Uecker, P. Reiche, Y. B. Chen, X. Q. Pan, V. Gopalan, L.-Q. Chen, D. G. Schlom, C. B. Eom, Enhancement ferroelectricity in strained BaTiO$_3$ thin films. *Science* **306,** 1005-1009 (2004).

39. E. J. Guo, K. Dorr, A. Herklotz, Strain controlled ferroelectric switching time of BiFeO$_3$ capacitors. *Appl. Phys. Lett.* **101,** 242908 (2012).

40. Y. Kim, A. Kumar, O. Ovchinnikov, S. Jesse, H. Han, D. Pantel, I. Vrejoiu, W. Lee, D. Hesse, M. Alex, S. V. Kalinin, First-order reversal curve probing of spatially resolved polarization switching dynamics in ferroelectric nanocapacitors. *ACS Nano* **6,** 491-500 (2012).

41. A. Stamm, D. J. Kim, H. Lu, C. W. Bark, C. B. Eom, A. Gruverman, Polarization relaxation kinetics in ultrathin ferroelectric capacitors. *Appl. Phys. Lett.* **102,** 092901 (2013).

42. Q. Yang, J. Hu, Y.-W. Fang, Y. Jia, R. Yang, S. Deng, Y. Lu, O. Dieguez, L. Fan, D. Zheng, X. Zhang, Y. Dong, Z. Luo, Z. Wang, H. Wang, M. Sui, X. Xing, J. Chen, J. Tian, L. Zhang, Ferroelectricity in latered bismuth oxide down to 1 nanometer. *Science* **379,** 1218-1224 (2023).
43. J. Gou, H. Bai, X. Zhang, Y. L. Huang, S. Duan, A. Ariando, S. A. Yang, L. Chen, Y. Lu, A. T. S. Wee, Two-demensional ferroelectricity in a single-element bismuth monolayer. *Nature* **617,** 67-72 (2023).
44. W. Li, X. Zhang, J. Yang, S. Zhou, C. Song, P. Cheng, Y.-Q. Zhang, B. Feng, Z. Wang, Y. Lu, K. Wu, L. Chen, Emergence of ferroelectricity in a nonferroelectric monolayer. *Nat. Commun.* **14,** 2757 (2023).
45. M. Weidemann, D. Werhahn, C. Mayer, S. Kläger, C. Ritter, P. Manuel, J. P. Attfield, Simon D. Kloß, High-pressure synthesis of Ruddlesden-Popper nitrides. *Nat. Chem.* **16,** 1723–1731 (2024).
46. H. Guo, Y. Huang, K. Jin, Q. Zhou, H. B. Lu, L. Liu, Y. Zhou, B. Cheng, and Z. Chen, Temperature effect on carrier transport characteristics in SrTiO$_{3-\delta}$//Si heterojunction, *Appl. Phys. Lett.* **86**, 123502 (2005).
47. H. F. Tian, H. X. Yang, H. R. Zhang, Y. Li, H. B. Lu, and J. Q. Li, Interface of epitaxial SrTiO$_3$ on silicon characterized by transmission electron microscopy, electron energy loss spectroscopy, and electron holography, *Phys. Rev. B* **73**, 075325 (2006).
48. Z. Li, X. Guo, H. B. Lu, Z. Zhang, D. Song, S. Cheng, M. Bosman, J. Zhu, Z. Dong, and W. Zhu, an epitaxial ferroelectric tunnel junction on silicon, *Adv. Mater.* **26**, 7185-7189 (2014).
49. P. Hohenberg, W. Kohn, Inhomogeneous electron gas. *Phys. Rev.* **136,** B864-B871 (1964).
50. W. Kohn, L. J. Sham, Self-consistent equations including exchange and correlation effects. *Phys. Rev.* **140,** A1133-A1138 (1963).
51. J. P. Perdew, A. Ruzsinszky, G. I. Csonka, O. A. Vydrov, G. E. Scuseria, L. A. Constantin, X. Zhou, K. Burke, Restoring the density-gradient expansion for exchange in solids and surfaces. *Phys. Rev. Lett.* **100,** 136406 (2008).
52. J. Heyd, G. E. Scuseria, Hybrid fuctionals based on a screened Coulomb potential. *J. Chem. Phys.* **118,** 8207-8215 (2003).
53. E. Kroumova, M. I. Aroyo, J. M. Perez-Mato, S. Ivantchev, J. M. Igartu, H. Wondratschek, PSEUDO: a program for a pseudo-symmetry search. *J. Appl. Cryst.* **34,** 783-784 (2001).
54. C. Capillas, M. I. Aroyo, J. M. Perez-Mato, Methods for pseudosymmetry evaluation: a comparison between the atomic displacements and eletron density approaches. *Z. Krist.* **220,** 691 (2005)
55. C. Capillas, E. S. Tasci, G. d. l. Flor, D. Orobengoa, J. M. Perez-Mato, M. I. Aroyo, A computer tool at the Bilbao crystallographic server to detect and characterize pseudosymmetry. *Z. Krist.* **226,** 186 (2011).
56. D. Sheppard, P. Xiao, W. Chemelewski, D. D. Johnson, G. Henkelman, A generalized solid-state nudged elastic band method. *J. Chem. Phys.* **136,** 074103 (2012).
57. D. Sheppard, G. Henkelman, Letter to the editor paths to which the nudged elastic band converges. *J. Comp. Chem.* **32,** 1769-1771 (2011).
58. A. N. Kolmogorov, *Izv. Akad. Nauk SSSR, Ser. Math.* **3**, 355 (1937).
59. M. Avrami, *J. Chem. Phys.* **8**, 212 (1940).

**Acknowledgements**

We thank Y. Fu, C. Zhou, and X. L. Zhang for preparation of nitride targets at the high-P Lab of SUSTech, Dr. X. K. Yao and Dr. X. Y. Wang for assistant PFM measurements at IOP-CAS. **Funding:** This work was supported by the Beijing Natural Science Foundation (Grant No. JQ24002 to E.J.G., JQ24011 to Q.L.), the National Key Basic Research Program of China (Grant Nos. 2020YFA0309100 to E.J.G., Q.L., and L.F.W.), the National Natural Science Foundation of China (Grant Nos. U22A20263 to E.J.G., 52250308 to E.J.G., 12474013 to S.M.W., 12304158 to Q.J., and 12174175 to L.F.W.), , the CAS Project for Young Scientists in Basic Research (Grant No. YSBR-084 to E.J.G.), the CAS Youth Interdisciplinary Team, the Special Research assistant, the CAS Strategic Priority Research Program (B) (Grant No. XDB33030200 to K.J.J. and E.J.G.), the Guangdong Basic and Applied Basic Research Foundation (Grant No. 2022B1515120014 to E.J.G.), the Guangdong-Hong Kong-Macao Joint Laboratory for Neutron Scattering Science and Technology, the China Postdoctoral Science Foundation (Grant No. 2022M723353 to E.J.G.), and the International Young Scientist Fellowship of IOP-CAS to S.C. XPS experiments were performed at IOP-CAS via a user proposal. The synchrotron-based XRD was conducted at the beamline 1W1A of the Beijing Synchrotron Radiation Facility (BSRF). **Author contributions:** E.J.G. initiated the research and supervised the work. The perovskite nitride targets were provided by S.W.; the perovskite nitride thin films were grown by S.C. and annealed using RTP with guidance from Q.J. and D.R.; TEM lamellas were fabricated with FIB milling and TEM experiments were performed by Q.H.Z. and L.G.; XPS measurements were performed by S.C., S.R.C., H.T.H., T.C., D.R., and Q.W.; PFM were performed by J.F., J.Z. and L.W.; S.X. and K.J.J worked on the SHG point measurements, and W.L. and Q.L. performed on SHG mapping measurements; X.Z., G.T., and H.JW. performed the first-principles calculations based on density functional theory. Z.G.C. performed the infrared spectroscopy measurements. C.G. and C.W. participated the discussions and provided suggestions during the manuscript preparation. S.C., X.Z., J.W.H., and E.J.G. wrote the manuscript with inputs from all authors. **Competing interests:** The authors declare that they have no competing interests. **Data and Materials availability:** All data needed to evaluate the conclusions in the paper are present in the paper and/or the Supplementary Materials.

**Table of contents for supplementary materials**

fig. S1. Band structures, as calculated with the HSE06 functional for $CeTaN_3$. The indirect band gaps are indicated.

fig. S2. The band gap evolution of $CeTaN_3$ as a function of biaxial strain.

fig. S3. The comparison of PDOS of all elements under different strain states.

fig. S4. Comparison of formation energy, cleavage energy, and surface energy for (001)-, and (011)-oriented $CeTaN_3$.

fig. S5. Powder X-ray diffraction $\theta$-$2\theta$ of $CeTaN_3$ ceramic target.

fig. S6. Infrared spectroscopy of crystalline $CeTaN_3$ with reference data from an amorphous $CeTaN_3$ and a $SrTiO_3$/Si substrate.

fig. S7. XRD measurements on amorphous and crystallized $CeTaN_3$ thin films.

fig. S8. STEM-HAADF image of a $CeTaN_3$ thin film grown on $SrTiO_3$ substrates.

fig. S9. Structural characterization of 3 nm-thick STO capped Si substrate.

fig. S10. XRD $\theta$-$2\theta$ of $CeTaN_3$ crystallization on STO/Si substrate using RTP.

fig. S11. Phi-scans for $CeTaN_3$ 022 and Si 011 reflections. Two sets of curves shift by 45 degrees.

fig. S12. Tetragonal structure of $CeTaN_3$ thin films confirmed by RSM.

fig. S13. X-ray photoelectron spectroscopy of $CeTaN_3$ thin films.

fig. S14. Determination of band gap of $CeTaN_3$.

fig. S15. ToF-SIMS analysis of $CeTaN_3$/$SrTiO_3$/$a$-$SiO_2$/Si.

fig. S16. High resolution STEM images for a $CeTaN_3$/$SrTiO_3$/a-$SiO_2$/Si sample.

fig. S17. Polarization analysis of $CeTaN_3$.

**Figures and figure captions**

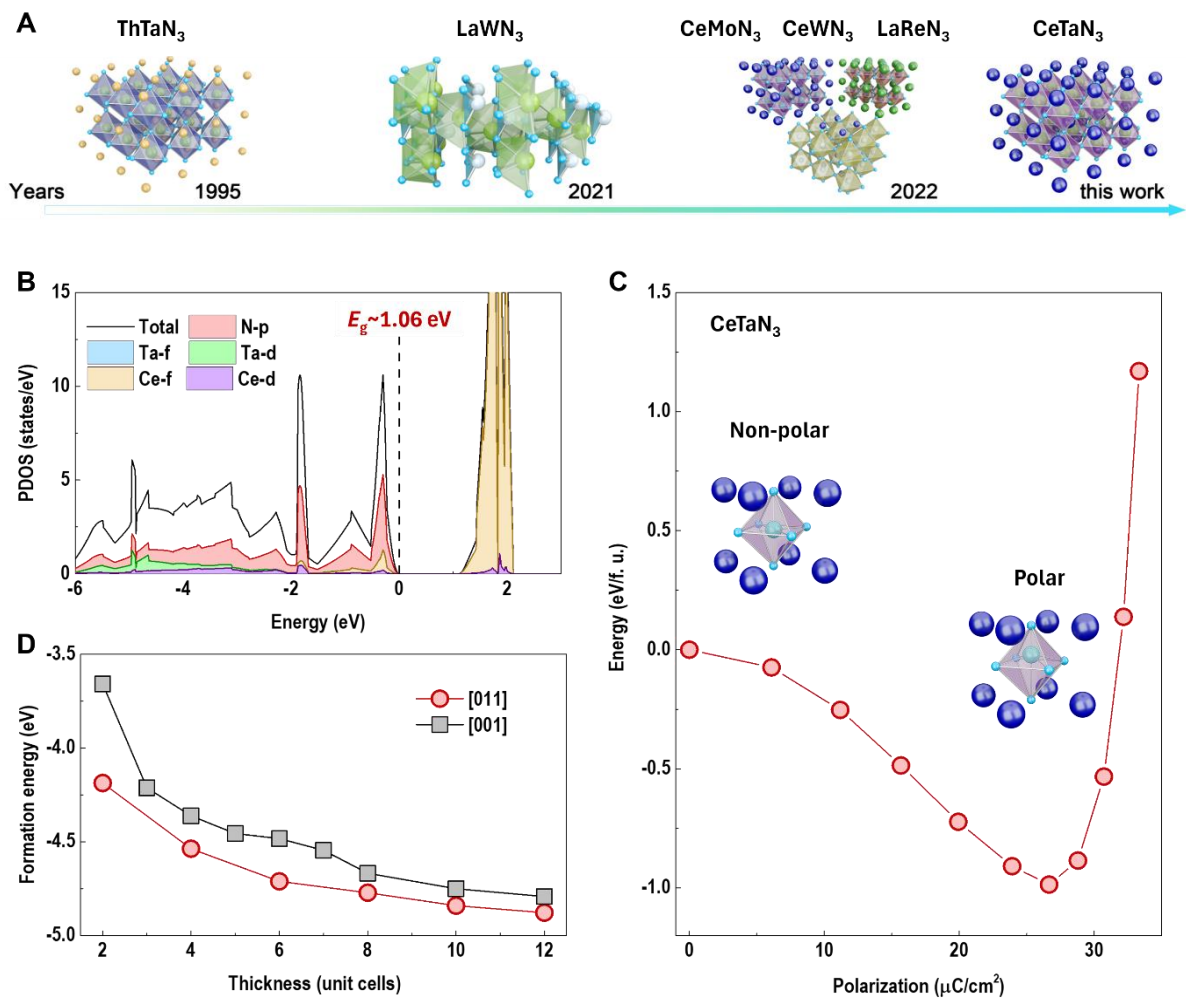

**Fig. 1. Theoretically predicted stable nitride perovskites.** (**A**) Theoretical and experimental biography of nitride perovskites (ABN$_3$) from first discovery of ThTaN$_3$ to this work. (**B**) The total and projected density of states (PDOS) of CeTaN$_3$ calculated using density function theory. The conduction bands are mainly occupied by Ce 4f electrons, whereas the valence bands are dominated by N 2$p$ electrons. (**C**) The free energy of strained CeTaN$_3$ thin films as a function of calculated ferroelectric polarization. The valley trend confirms that the CeTaN$_3$ thin films can be stabilized in a ferroelectric polar structure. (**D**) Comparison of formation energies of (001)- and (011)-oriented CeTaN$_3$ with different film thicknesses.

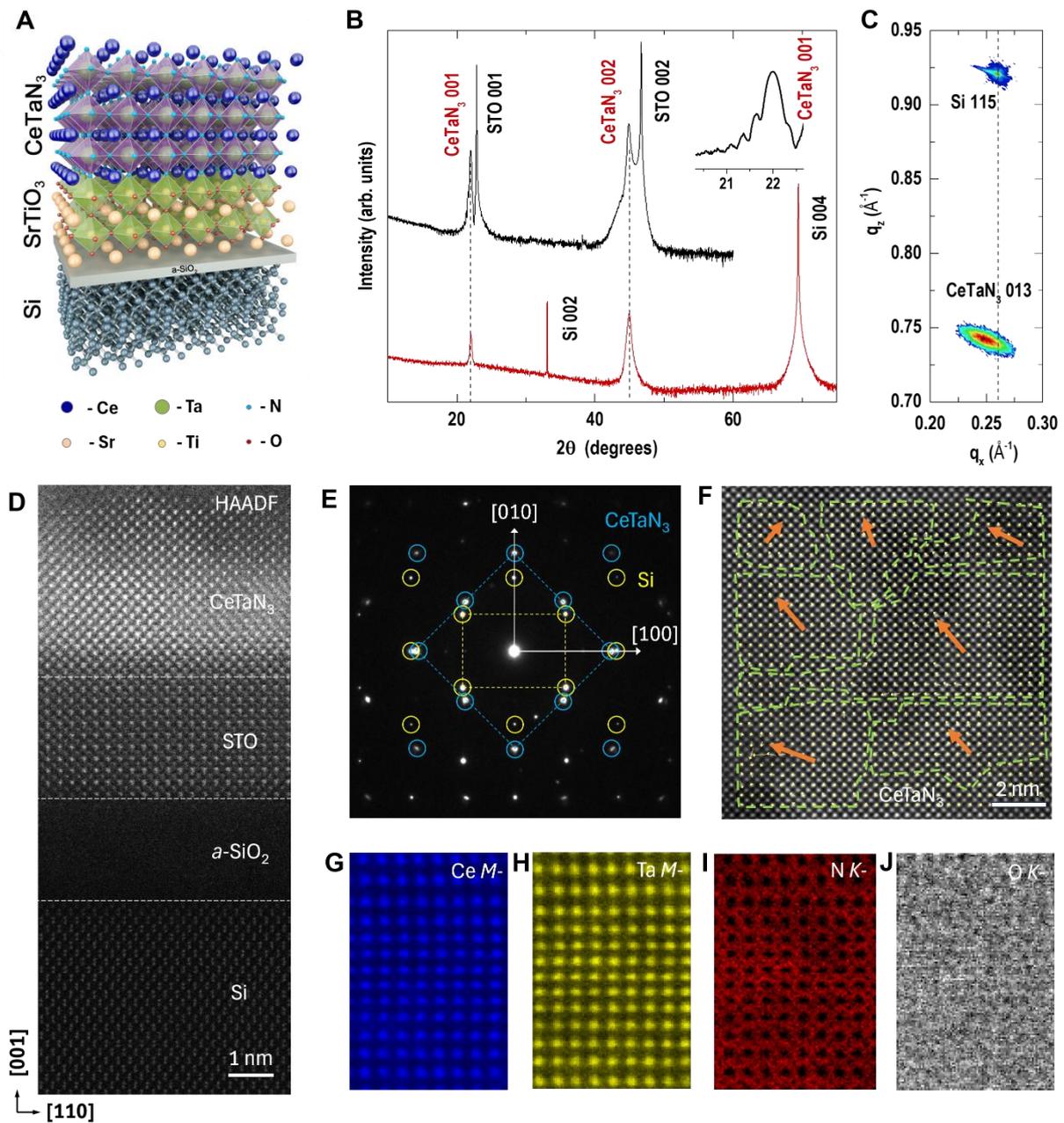

**Fig. 2. High-quality single-crystalline CeTaN$_3$ thin films.** (**A**) Schematics of CeTaN$_3$/SrTiO$_3$/$a$-SiO$_2$/Si substrates. An ultrathin amorphous Si was formed under the growth conditions. The unit cells of CeTaN$_3$, SrTiO$_3$, and Si were identified. (**B**) X-ray diffraction $\theta$-$2\theta$ scans of CeTaN$_3$ on SrTiO$_3$ and SrTiO$_3$/Si substrates containing an inset figure of zoom-in CeTaN$_3$ 001 diffraction peak with clear Laue oscillations. (**C**) Reciprocal space mapping of CeTaN$_3$ 013 around the reflection of Si 115. (**D**) High-resolution HAADF-STEM image and (**E**) corresponding SAED pattern of CeTaN$_3$/SrTiO$_3$/$a$-SiO$_2$/Si. Blue and yellow circles identify the electron diffractions from CeTaN$_3$ and Si, respectively. (**F**) HAADF-STEM image of a representative region from single-crystalline CeTaN$_3$. The yellow arrows indicate the cation displacement vectors of central Ta ions in each unit cell. (**G** to **J**) Compositional EELS mapping taken simultaneously at Ce $M$-, Ta $M$-, N $K$-, and O $K$-edges, respectively.

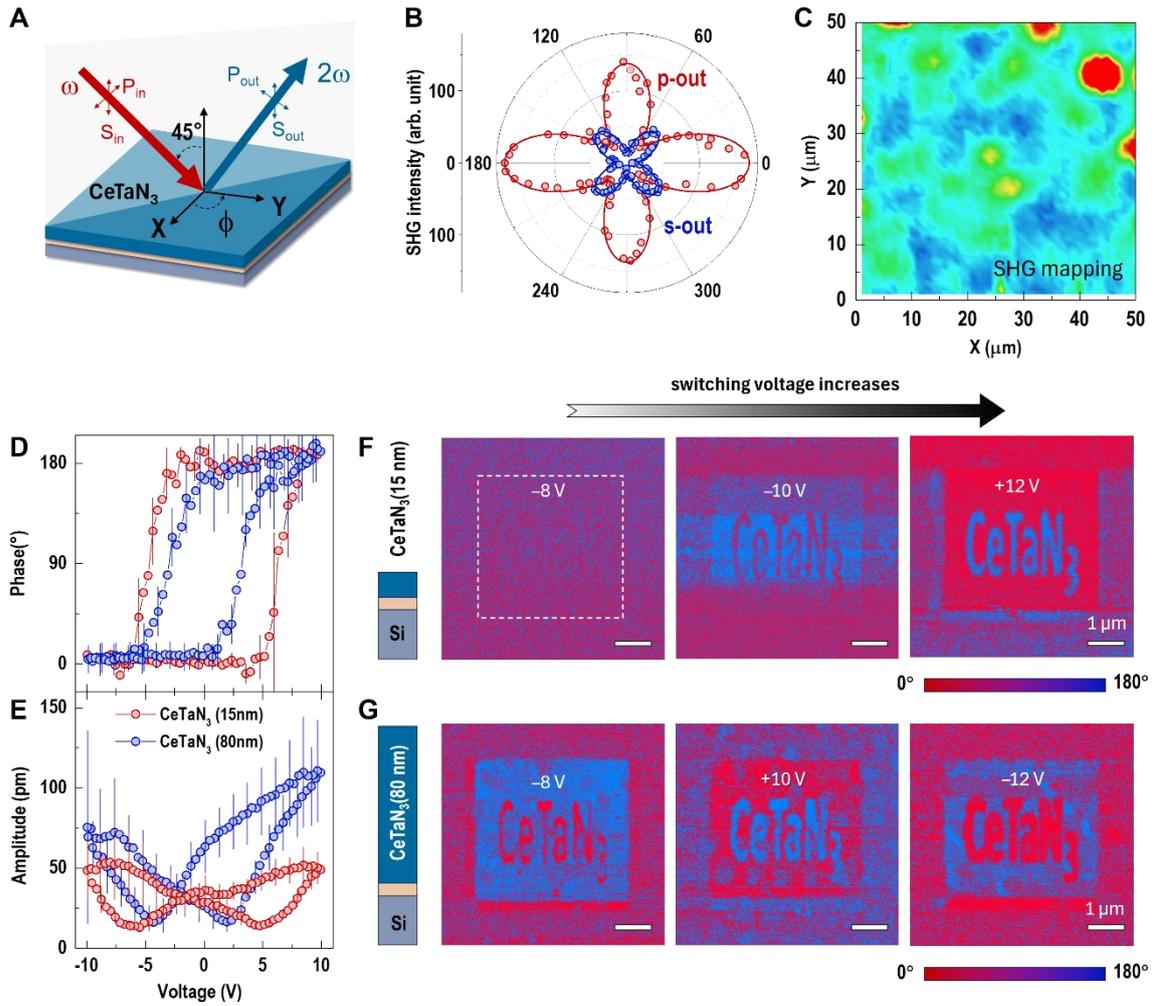

**Fig. 3. Switchable polar structure of CeTaN₃ thin films confirmed by SHG and PFM.** (**A**) Schematic of SHG setup. (**B**) SHG results of 2ω responses. (**C**) SHG mapping conducted over an area of 50 × 50 µm² on an CeTaN₃ thin film. (**D**) PFM phase and (**E**) PFM amplitude hysteresis loops from a 15-nm-thick and an 80-nm-thick CeTaN₃ thin films, respectively. PFM phase images of (**F**) a 15-nm-thick and (**G**) an 80-nm-thick CeTaN₃ thin films with progressively increasing switching voltages. A square area was switched uniformly and then the characters "CeTaN₃" were written with opposite tip voltages.

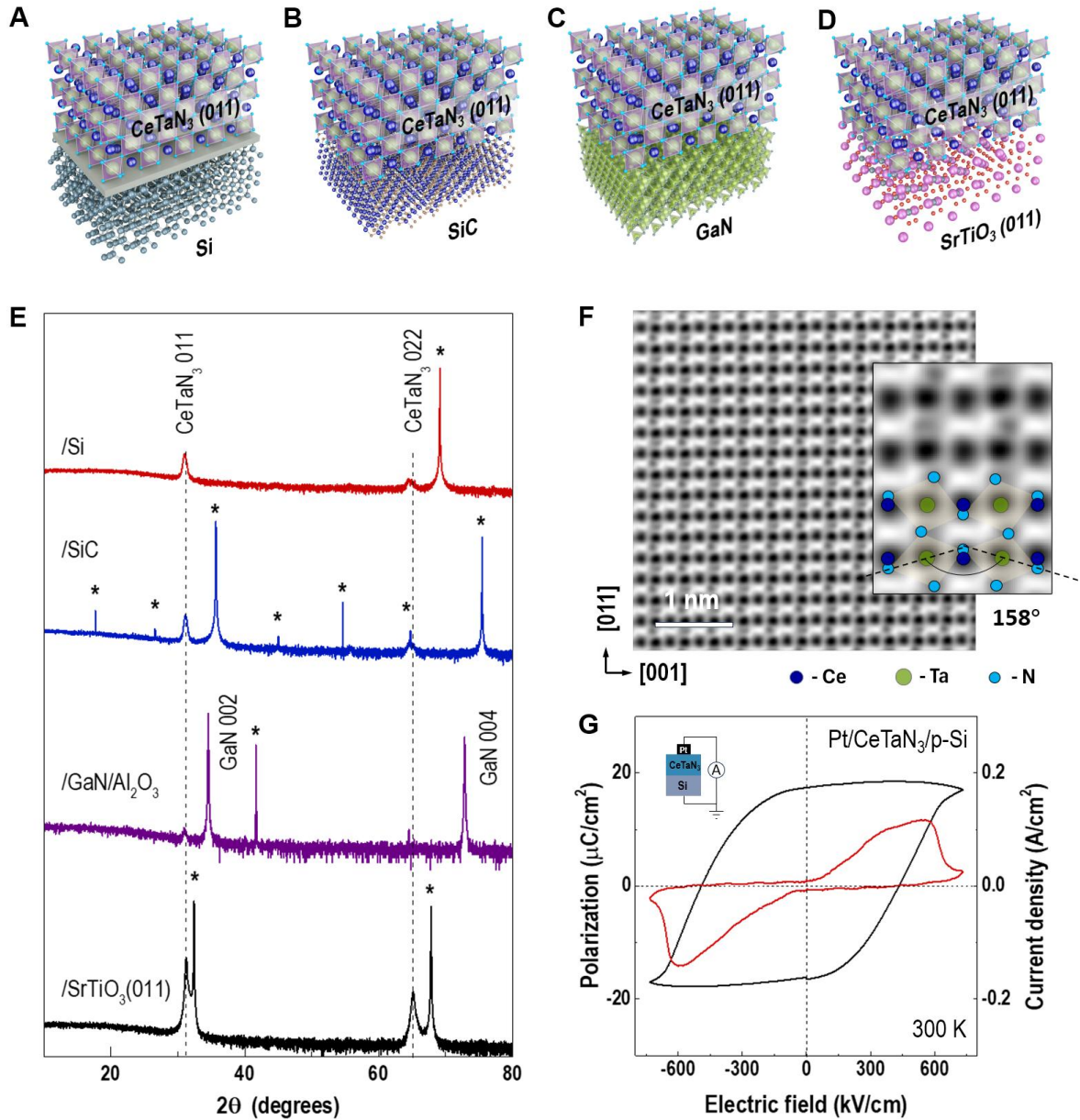

**Fig. 4. Naturally formed (011)-oriented CeTaN₃ thin films on wide-range semiconductors.** (**A** to **D**) Schematics of (011)-oriented $CeTaN_3$ thin films grown on Si, SiC, GaN/Al$_2$O$_3$, and (011)-oriented SrTiO$_3$ substrates, respectively. (**E**) X-ray diffraction $\theta$-$2\theta$ scans of $CeTaN_3$ thin films on various substrates (indicated with "*"). Dashed lines indicate the positions of (0$ll$) peaks from $CeTaN_3$ thin films. (**F**) High-magnified STEM-ABF image from a (011)-oriented $CeTaN_3$ thin film grown on Si. Inside shows a representative tilted $TaN_6$ octahedral with an averaged Ta-N-Ta bond angle of ~ 158°. (**G**) Ferroelectric hysteresis loops and switching currents of a Pt/CeTaN₃/*p*-Si capacitor. The measurements were conducted at room temperature. Inset shows the schematic of measurement setup.

Supplementary Materials for

# A single-phase epitaxially grown ferroelectric perovskite nitride

Songhee Choi, Qiao Jin, Xian Zi, Dongke Rong, Jie Fang, Jinfeng Zhang, Qinghua Zhang, Wei Li, Shuai Xu, Shengru Chen, Haitao Hong, Cui Ting, Qianying Wang, Gang Tang, Chen Ge, Can Wang, Zhiguo Chen, Lin Gu, Qian Li, Lingfei Wang, Shanmin Wang, Jiawang Hong, Kuijuan Jin, and Er-Jia Guo

*Corresponding author Emails: kjjin@iphy.ac.cn, wangsm@sustech.edu.cn, hongjw@bit.edu.cn, and ejguo@iphy.ac.cn

**This PDF file includes:**

    figs. S1 to S30

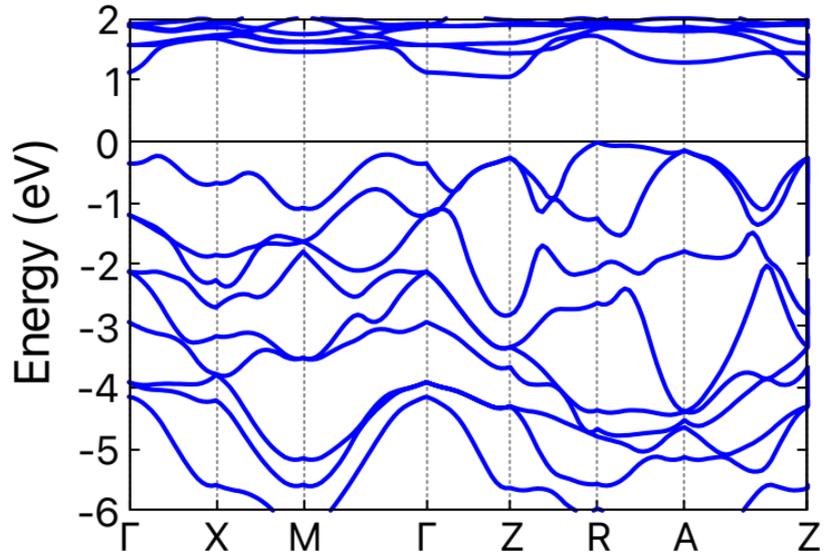

**fig. S1. Band structures, as calculated with the HSE06 functional for CeTaN$_3$.** The indirect band gaps are indicated. The zero of energy is set at the valence-band maximum.

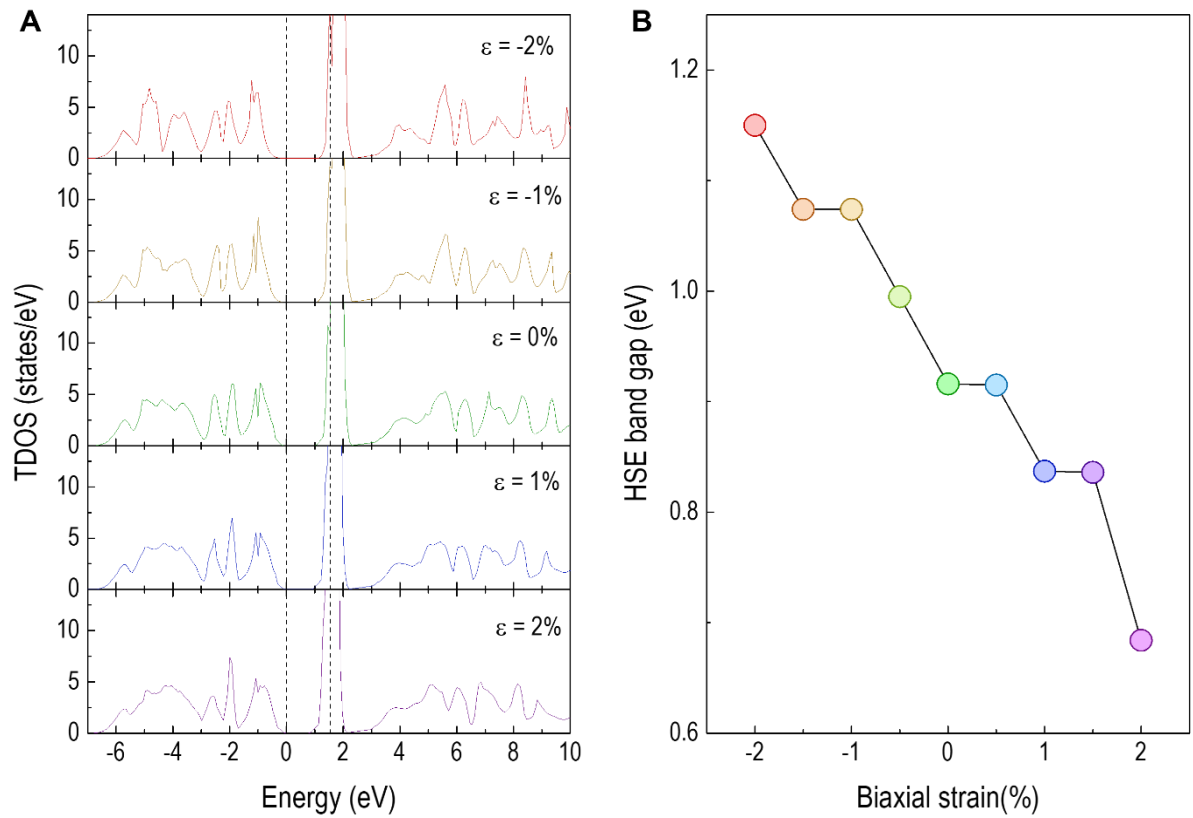

**fig. S2. The band gap evolution of CeTaN$_3$ as a function of biaxial strain.** (**A**) Total DOS of CeTaN3 with biaxial strain ranging from –2% to 2%. Increasing epitaxial strain reduces the band gap, which is mainly caused by the downward movement of conduction band. (**B**) HSE bandgap evolution as a function of biaxial strain.

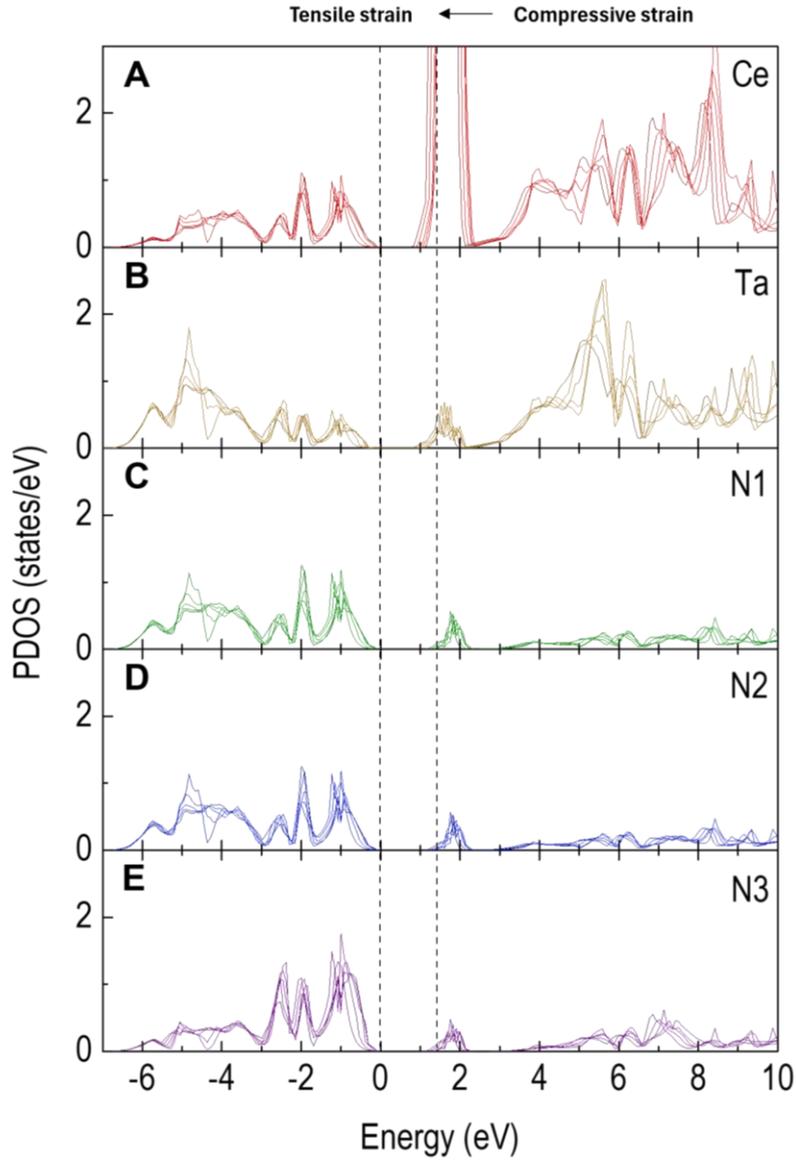

**fig. S3. The comparison of PDOS of all elements under different strain states.** PDOS of (**A**) Ce, (**B**) Ta, (**C**) N1, (**D**) N2, and (**E**) N3. Biaxial strain produces an additional energy shift to all atoms, leading to a smaller band gap under tensile strain and a larger band gap under compressive strain. Both tensile and compressive strain causes conduction band downward, while only compressive strain causes the valence band upward.

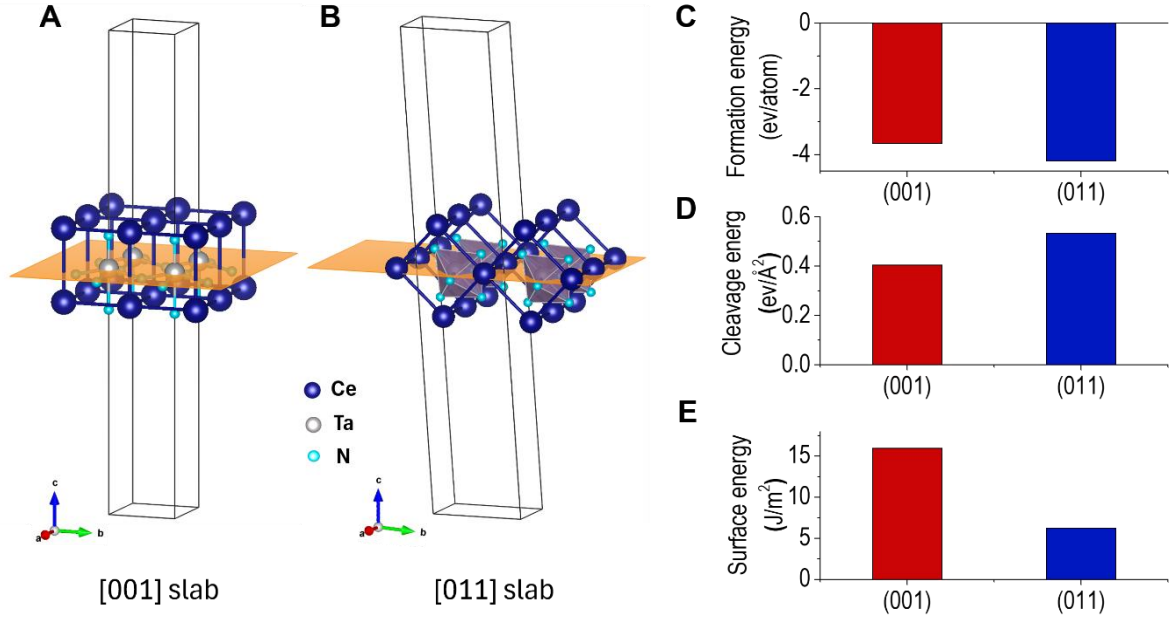

**fig. S4. Comparison of formation energy, cleavage energy, and surface energy for (001)-, and (011)-oriented CeTaN₃.** We can calculate the formation energy Ef by $(E_{slab} - n(E_{Ce} + E_{Ta} + 3*E_N))/N_{atom}$, the cleavage energy Ecle by $(E_{top}^{unrel} + E_{bottom}^{unrel} - E_{bulk})/2S$, and the surface energy Esurf by (Eslab – nEbulk)/A, where Eslab is the total energy of the relaxed CeTaN₃, Ebulk is the total energy of CeTaN₃ bulk with the same crystal structure. $E_{Ce}$, $E_{Ta}$, and $E_N$ are total energy of single atom of Ce, Ta, and N, respectively. $E_{top}^{unrel}$ is the unrelaxed energy of the top layer after cleavage, $E_{bottom}^{unrel}$ is the unrelaxed energy of the bottom layer after cleavage. All energies can be calculated using first-principles calculations based on density functional theory.

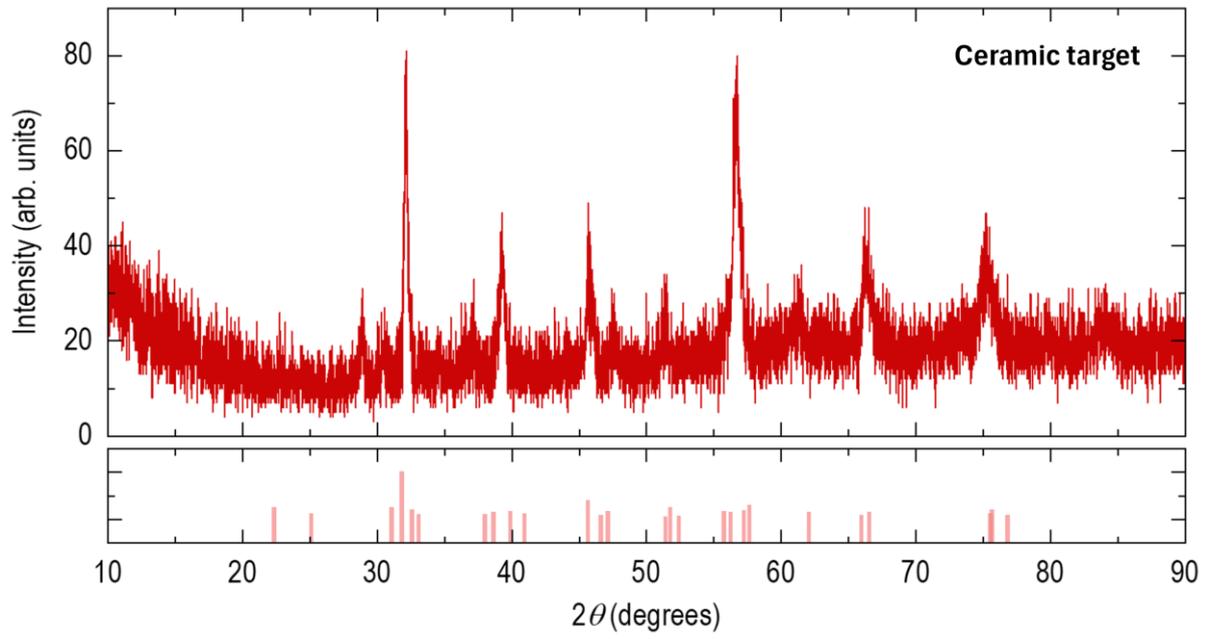

fig. S5. Powder X-ray diffraction $\theta$-$2\theta$ of CeTaN$_3$ ceramic target.

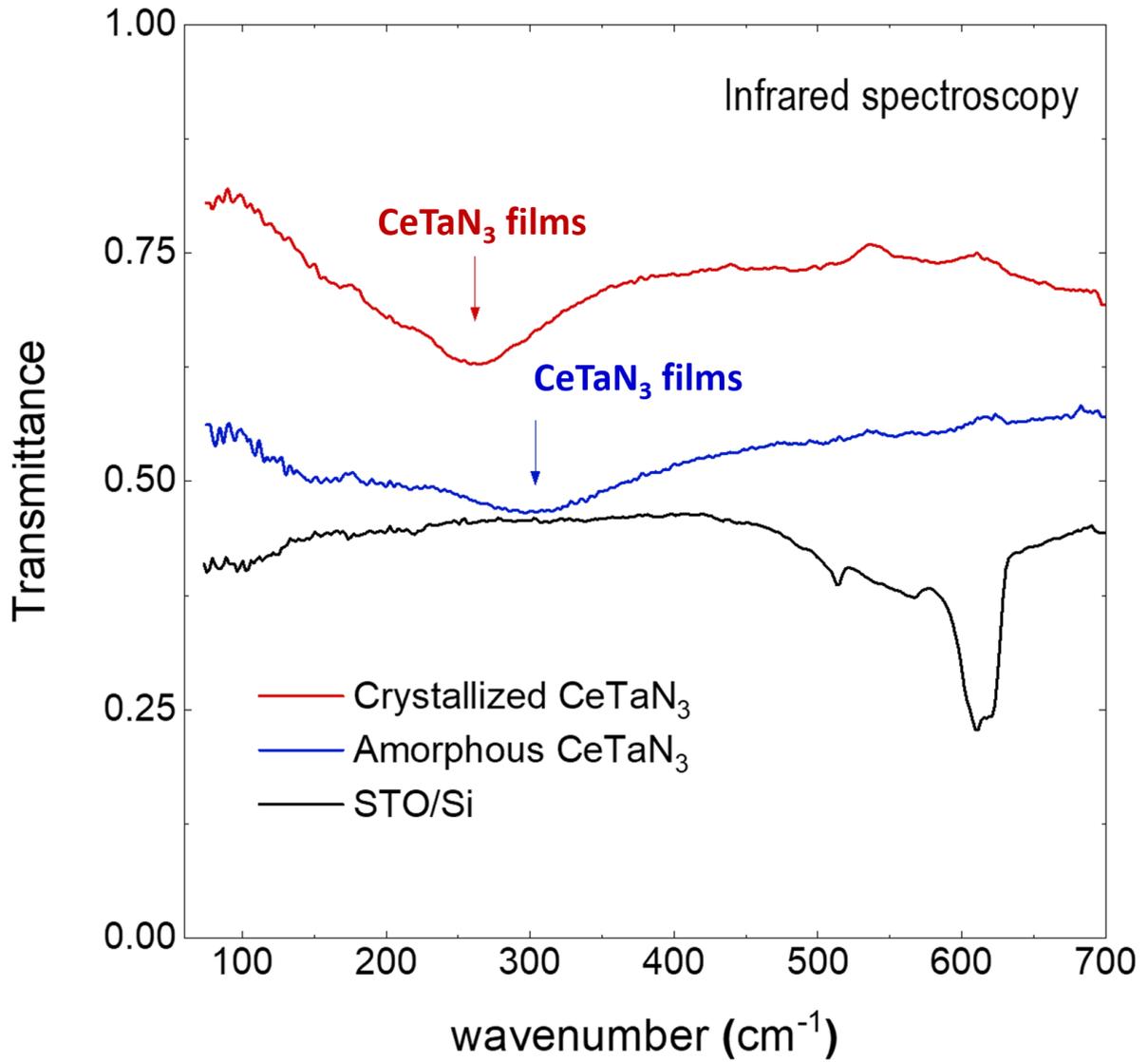

**fig. S6. Infrared spectroscopy of crystalline CeTaN$_3$ with reference data from an amorphous CeTaN$_3$ and a SrTiO$_3$/Si substrate.** Arrow lines mark the presence of CeTaN$_3$, in sharp difference with pure SrTiO$_3$/Si substrates.

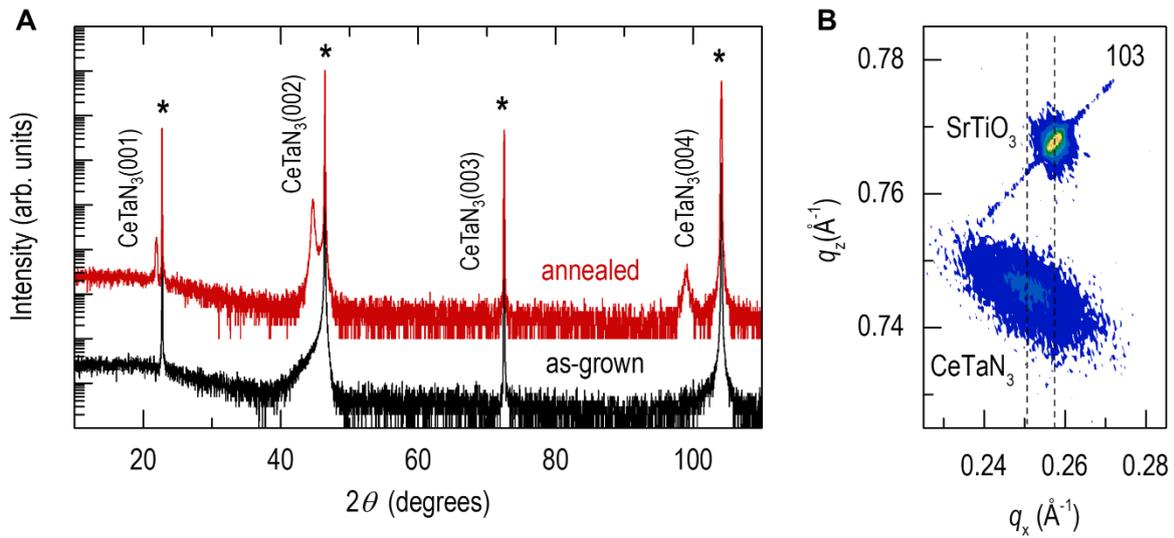

**fig. S7. XRD measurements on amorphous and crystallized CeTaN$_3$ thin films on (001)-oriented SrTiO$_3$ substrates using rapid thermal process (RTP).** (**A**) XRD θ-2θ scans of amorphous CeTaN$_3$ and crystalline CeTaN$_3$ before and after RTP, respectively. RTP was performed at the temperature of 800°C in ammonia (NH$_3$) conditions with 10,000 Pa for a 1 hour with heating rate of 200 °C/min. (**B**) RSM of CeTaN$_3$ around 103 peaks of SrTiO$_3$, yielding the in-plane and out-of-plane lattice constants of CTN films are 3.99 and 4.02 Å, respectively.

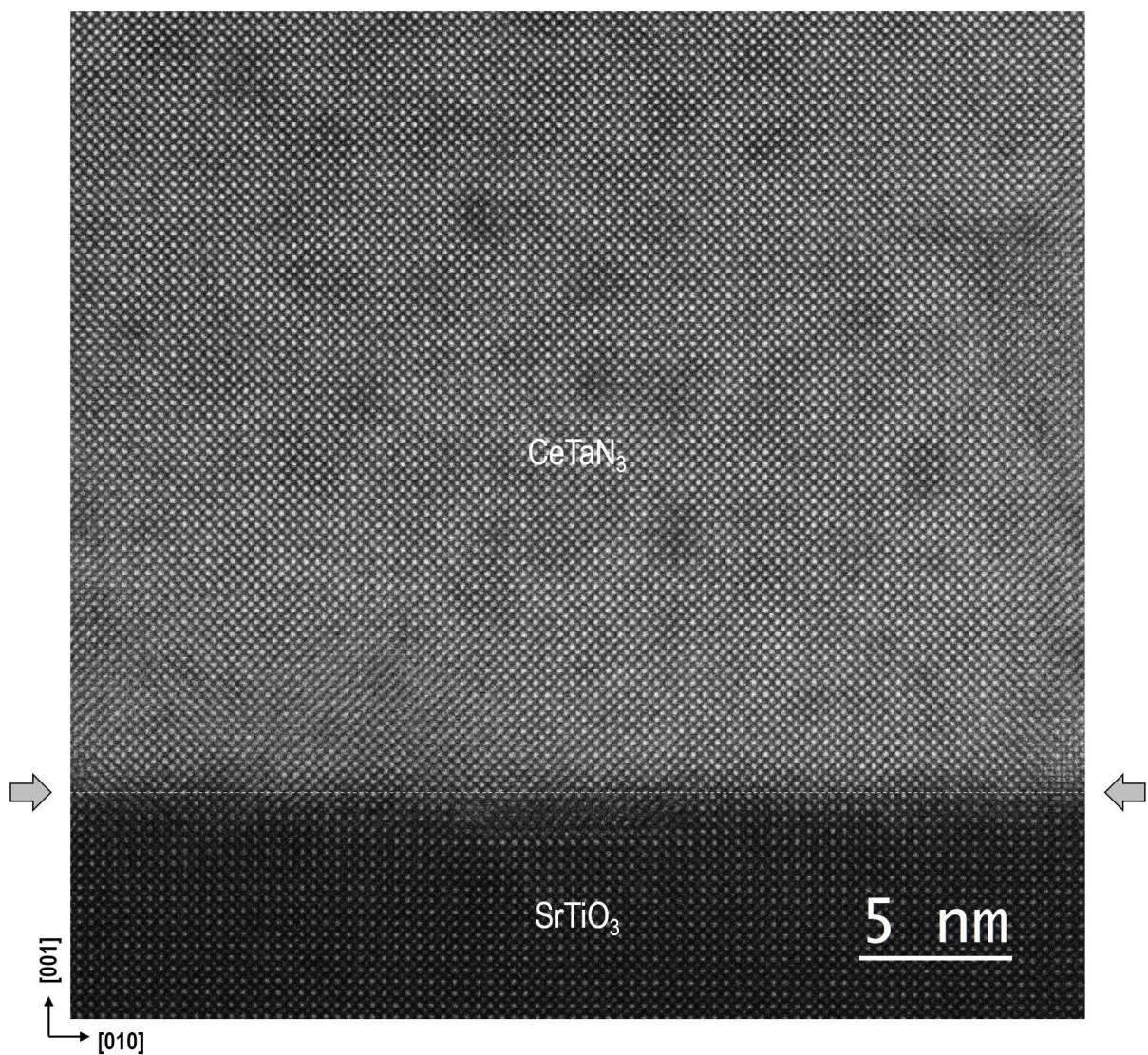

**fig. S8. STEM-HAADF image of a CeTaN$_3$ thin film grown on SrTiO$_3$ substrates.** Dashed line and arrows indicate the interfaces between CeTaN$_3$ and SrTiO$_3$. The sharp interfaces and well-ordered atoms indicate that the CeTaN$_3$ thin films have high crystallinity.

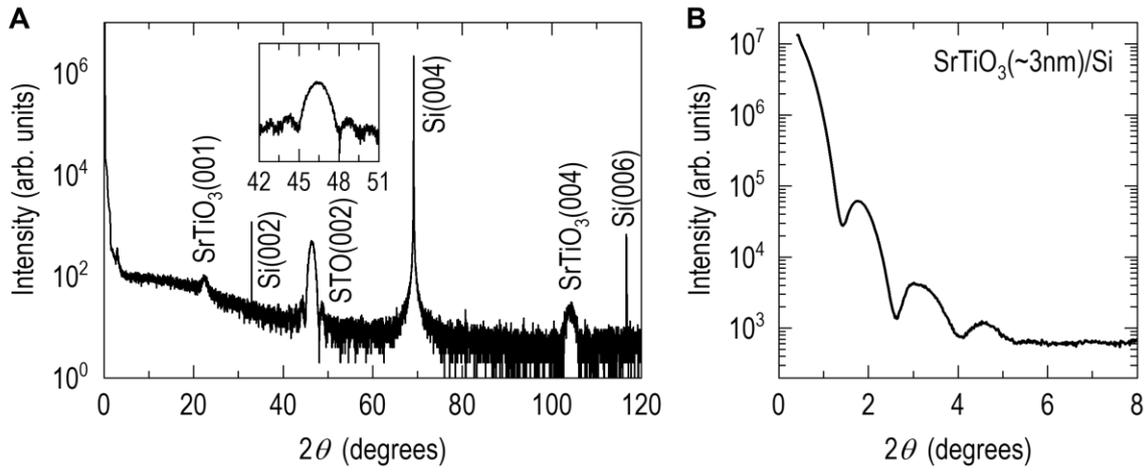

**fig. S9. Structural characterization of 3 nm-thick STO capped Si substrate.** (**A**) XRD θ-2θ scan exhibits the epitaxial growth of STO on Si without any impurity peaks. The inset shows the detail of the STO 002 peak, revealing the Kiessig fringes. (**B**) X-ray reflection (XRR) shows Fourier oscillations, confirming the flat surface and about 3nm thickness of the STO capping layer.

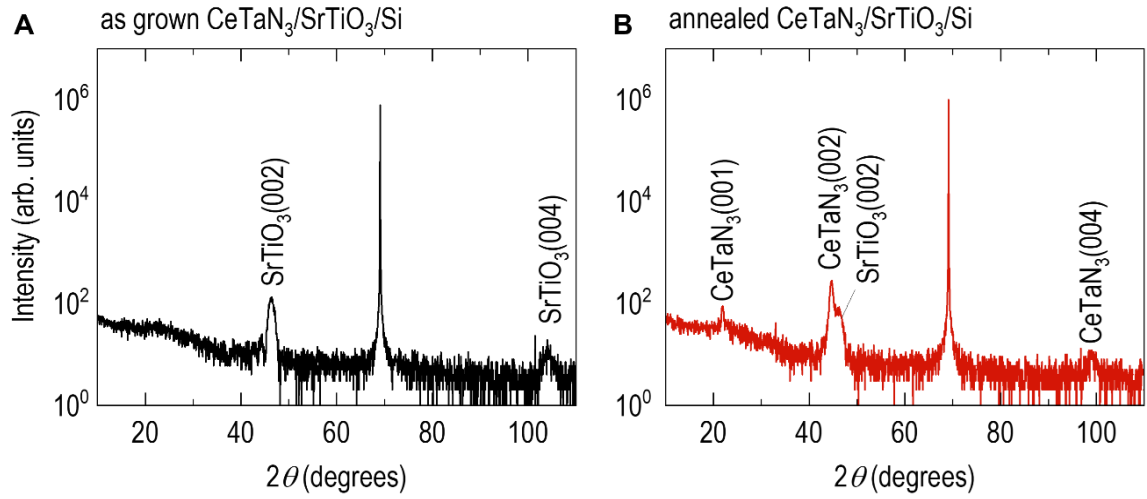

**fig. S10. XRD $\theta$-2$\theta$ of CeTaN$_3$ crystallization on STO/Si substrate using RTP.** (**A**) Before RTP, XRD result only shows the STO peaks without any peaks from CeTaN$_3$ films, suggesting that the CeTaN$_3$ is amorphous. (**B**) We find the 00$l$ reflections of CeTaN$_3$ films, indicating that the CeTaN$_3$ films are crystallized after RTP.

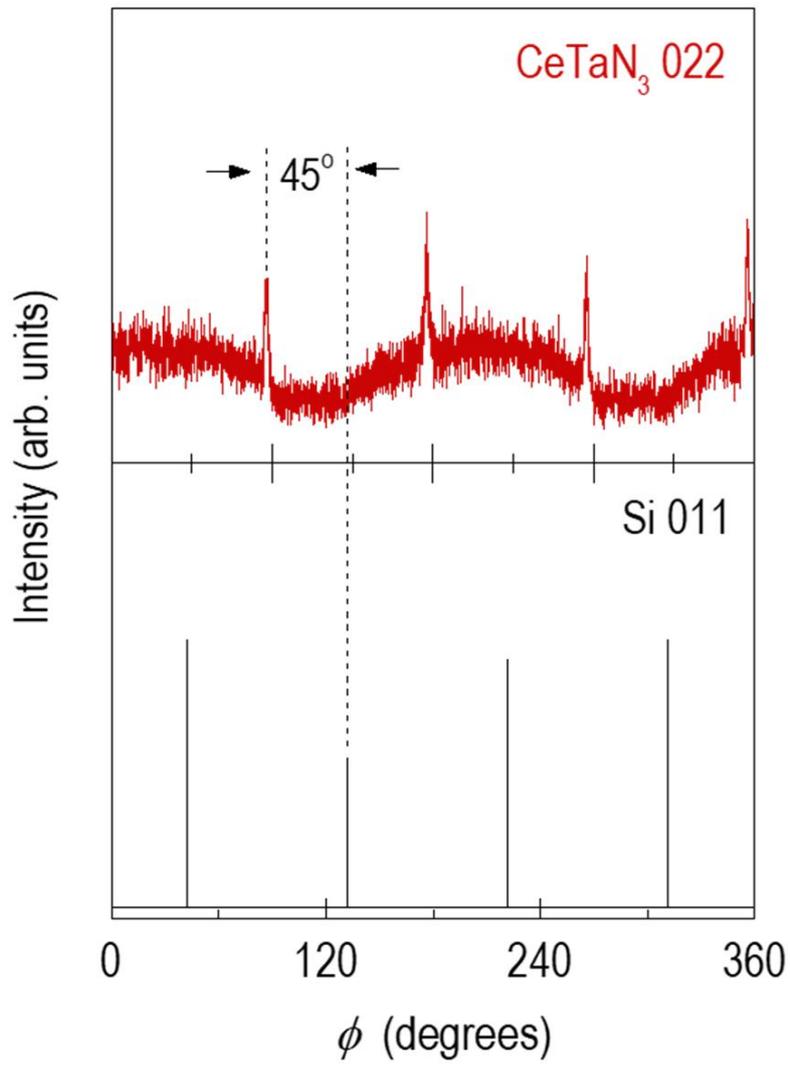

**fig. S11. Phi-scans for CeTaN$_3$ 022 and Si 011 reflections.** Two sets of curves shift by 45 degrees.

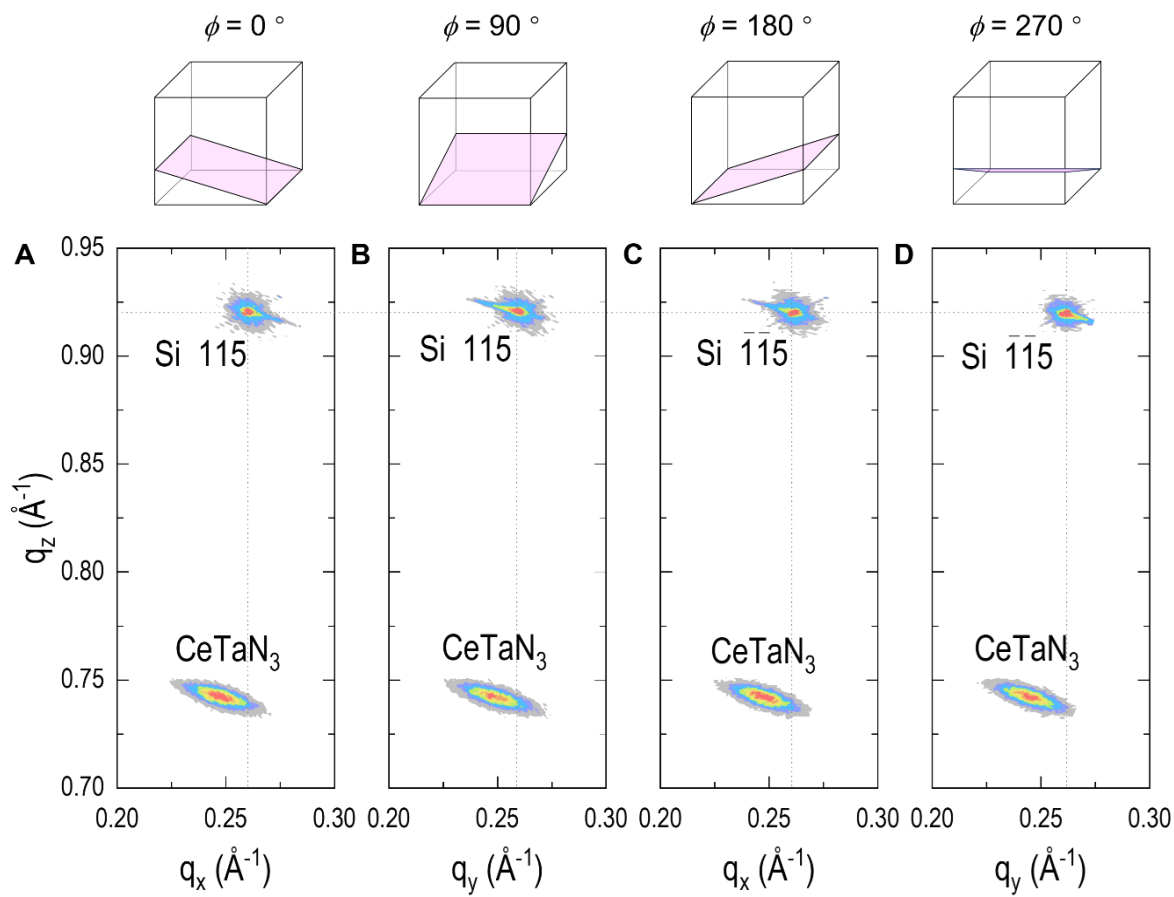

**fig. S12. Tetragonal structure of CeTaN$_3$ thin films confirmed by RSM.** The diffraction peaks, associated with the indices 103, 013, $\bar{1}$03, and 0$\bar{1}$3 reflections of CeTaN$_3$, reveal consistent positioning of q$_x$ and q$_y$. The phi-dependent RSM results suggest that the CeTaN$_3$ thin film exhibits a tetragonal structure. The average in-plane lattice parameter is calculated to be *a* = *b* = 4.03 Å, while the averaged out-of-plane lattice parameter is *c* = 4.09 Å.
32

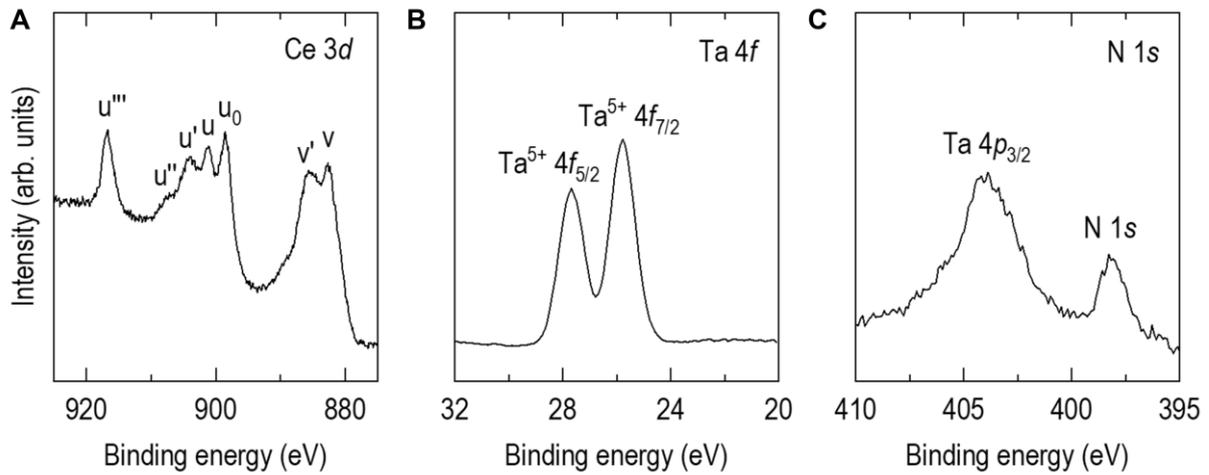

**fig. S13. X-ray photoelectron spectroscopy of CeTaN$_3$ thin films.** Ce 3$d$, Ta 4$d$, and N 1$s$, respectively. (**A**) Ce represents two oxidation states of Ce$^{3+}$ and Ce$^{4+}$. The peaks of u' (901.3 eV), u$_0$ (898.5 eV), and v' (885.5 eV) correspond to the positions of Ce$^{3+}$ determined from Ce$_2$O$_3$, while peaks of u''' (916.8 eV), u'' (907.2 eV), u (901.1 eV) and v (882.7 eV) match Ce$^{4+}$ determined from CeO$_2$. The presence of two oxidization states may attribute to the inevitable nitrogen vacancies in the surface layers of CeTaN$_3$ thin films. (**B**) The result of Ta spectroscopy reveals a single oxidation state of Ta$^{4+}$. Ta$^{5+}$ 4$f_{5/2}$ is located at 27.7 eV, and Ta$^{5+}$ 4$f_{7/2}$ is at 25.8 eV. (**C**) N shows a distinct single peak at 398.1 eV, nearby where a peak corresponding to Ta 4$p_{3/2}$ was observed at 403.9 eV.

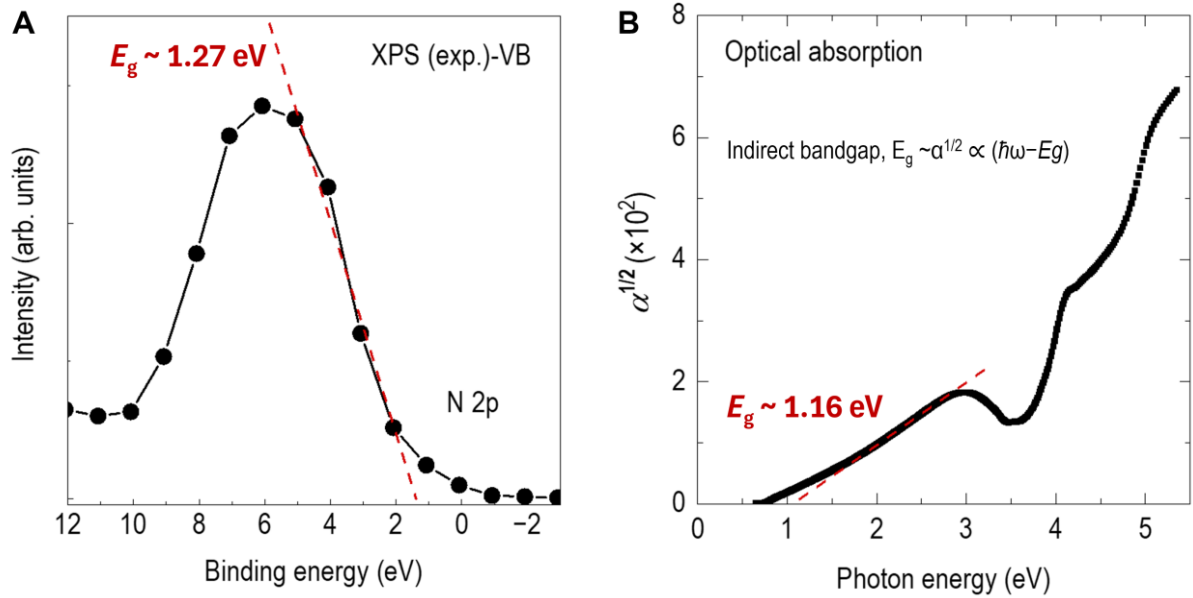

**fig. S14**. **Determination of band gap of CeTaN$_3$.** (**A**) XPS valence band (VB) spectra for CeTaN$_3$ thin films. The band gap estimated from VB spectra is approximately 1.27 eV. (**B**) Optical absorption spectra of a CeTaN$_3$ film on double-polished SrTiO$_3$ substrate. α represents the absorption coefficient. We plotted the α$^{1/2}$ against the photon energy ($\hbar\omega$). The linear relationship between α$^{1/2}$ and $\hbar\omega$ demonstrates that the CeTaN$_3$ is an indirect band gap semiconductor, consistent with theoretical calculations. The optical band gap determined from absorption spectra is approximately 1.16 eV.

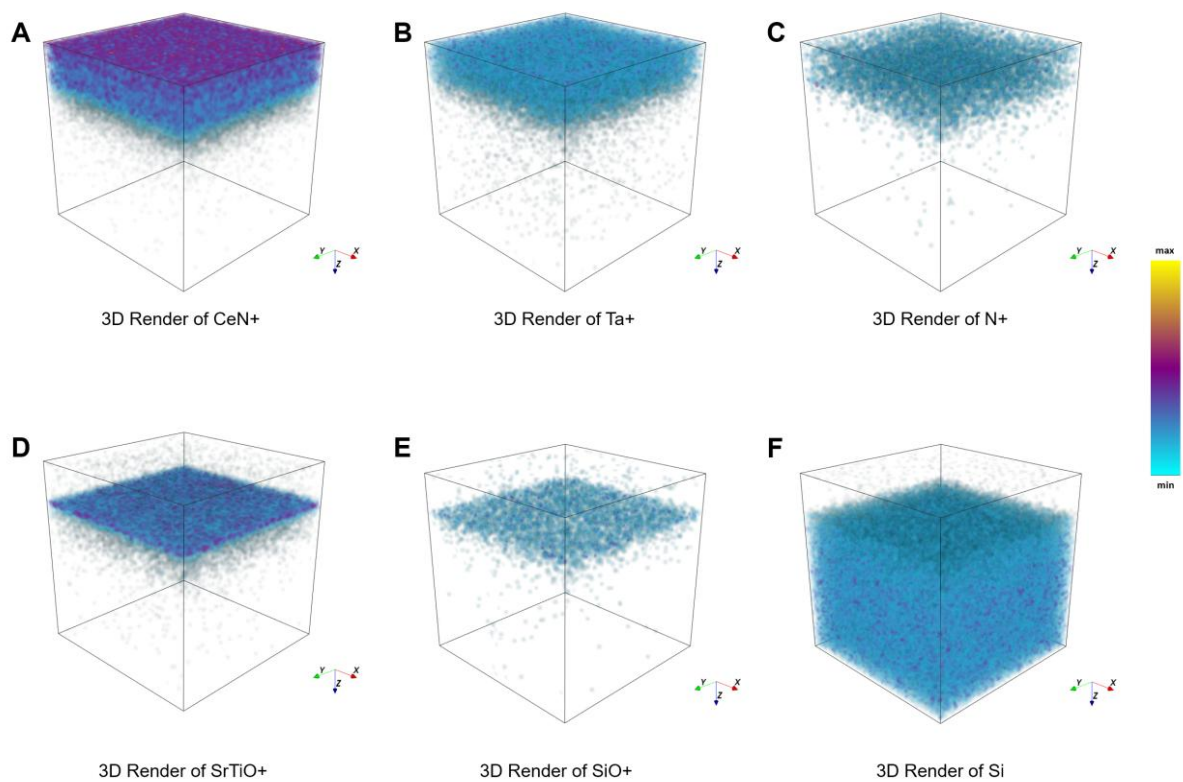

**fig. S15. ToF-SIMS analysis of CeTaN₃/SrTiO₃/*a*-SiO₂/Si.** (A–F) 3D renderings of CeN⁺, Ta⁺, N⁺, SrTiO⁺, SiO⁺, and Si signals, respectively. The SIMS results confirm the uniformity of all deposited layers and provide insight into the chemical distribution from the top to the bottom of the structure.

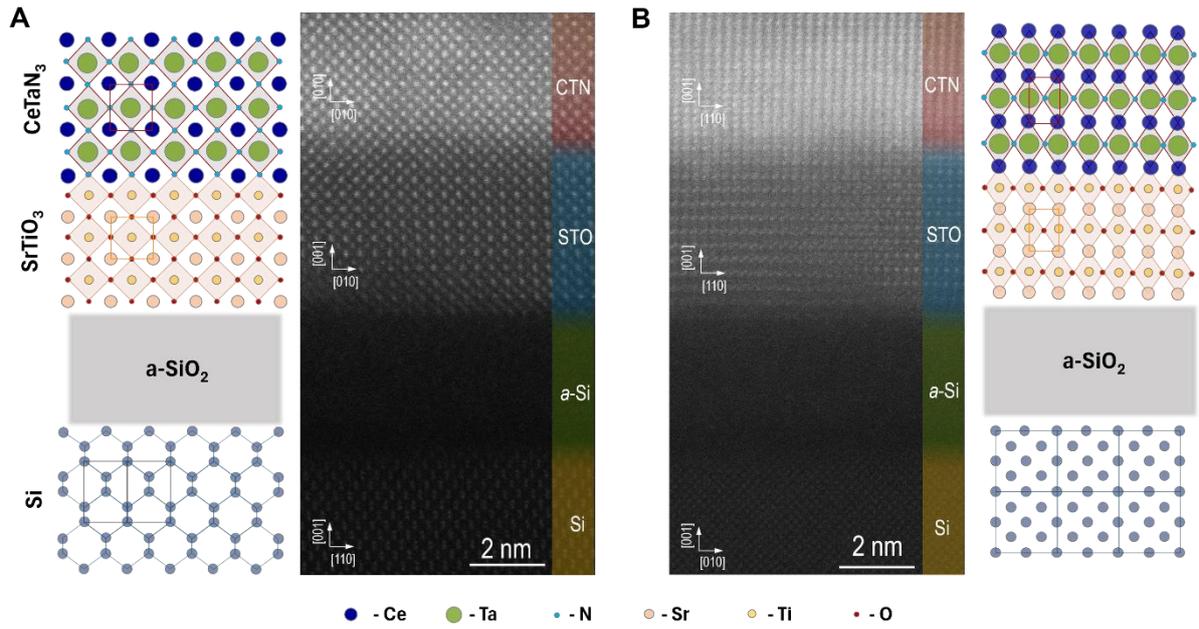

**fig. S16. High resolution STEM images for a CeTaN$_3$/SrTiO$_3$/*a*-SiO$_2$/Si sample.** (**A**) and (**B**) illustrate the STEM images viewed from Si [110] and Si [010] orientations, respectively.

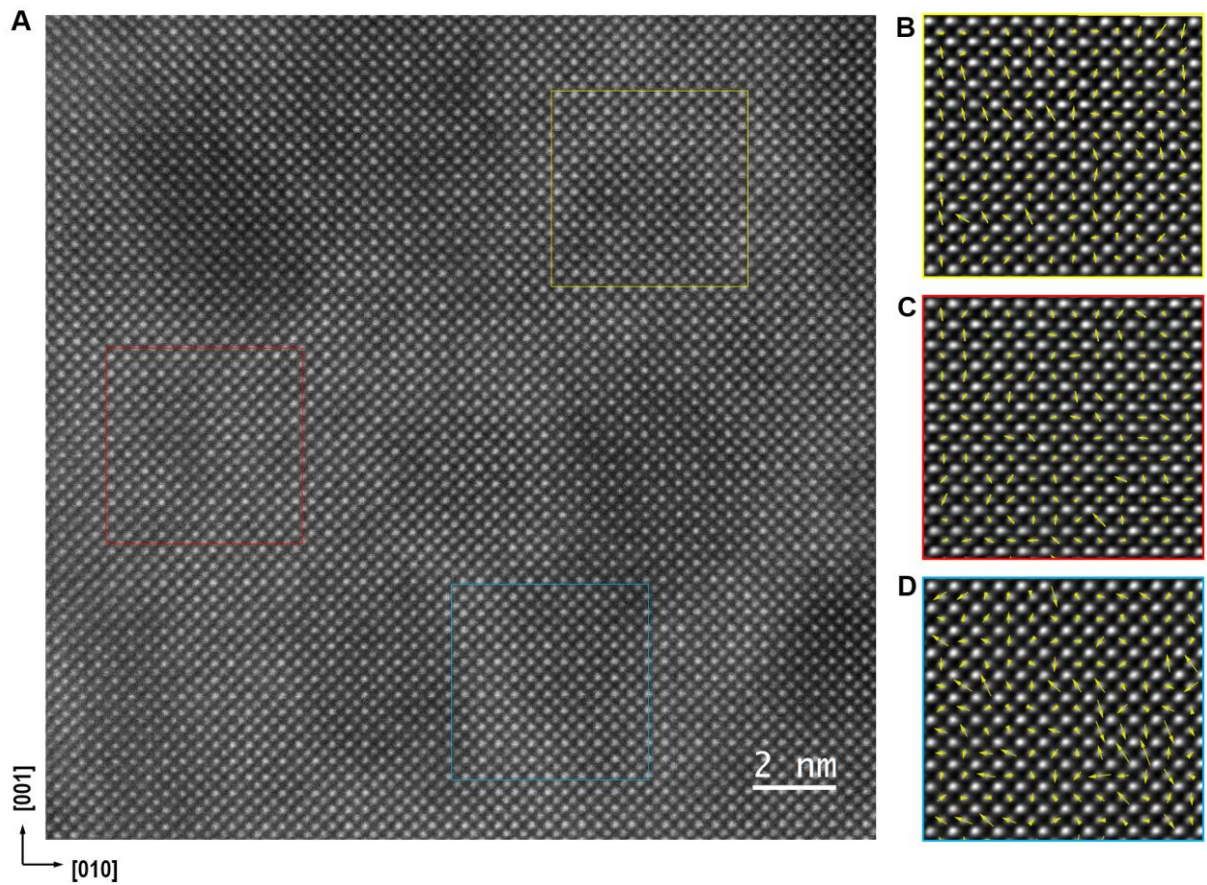

**fig. S17. Polarization analysis of CeTaN$_3$.** (**A**) A representative of zoom-in HAADF image of CeTaN$_3$. (**B** to **D**) The Ce atoms displacements were deduced from three representative TEM images. The arrows representing the motion of Ce atoms results to partial polarization, with most of these arrows pointing upward and some pointing to the left.

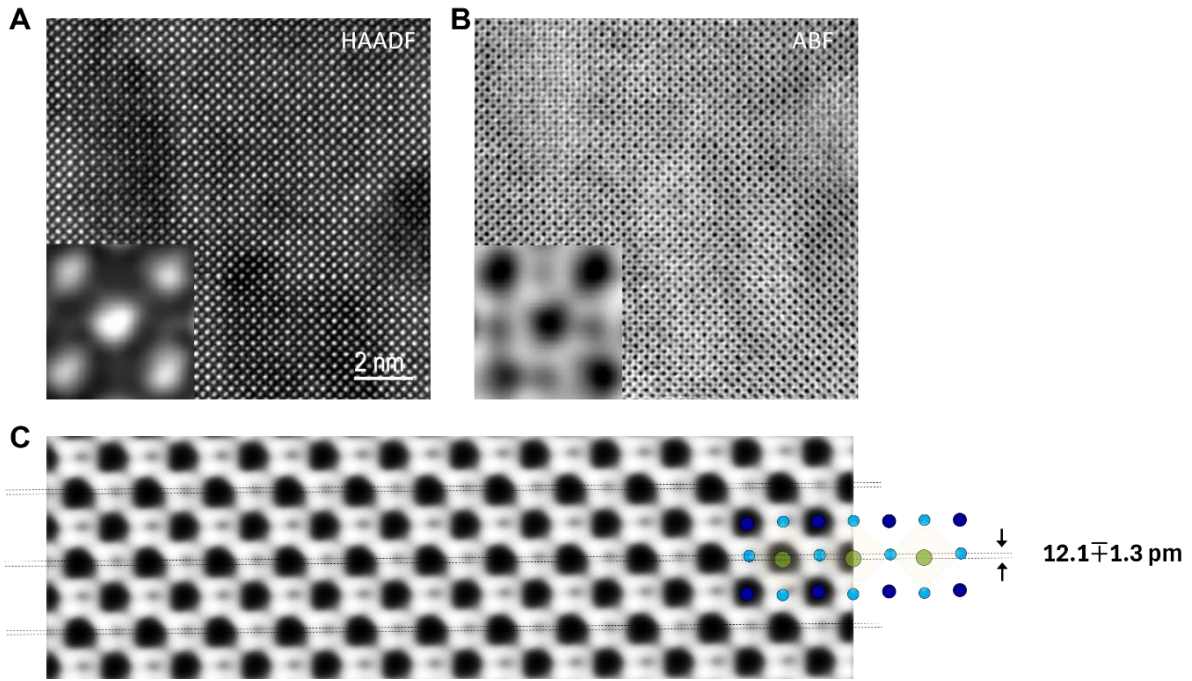

**fig. S18. STEM results of CeTaN$_3$ thin films.** To investigate atomic displacements intuitively related with ferroelectricity, **(A)** HAADF (high angle annular dark field) and **(B)** ABF (annular bright field) STEM images were recorded. **(C)** A ABF-STEM image of CeTaN$_3$, exhibiting schematic of non-centrosymmetric Ta atoms. The atomic displacement shifts from center position by 12.1 ± 1.3 pm.

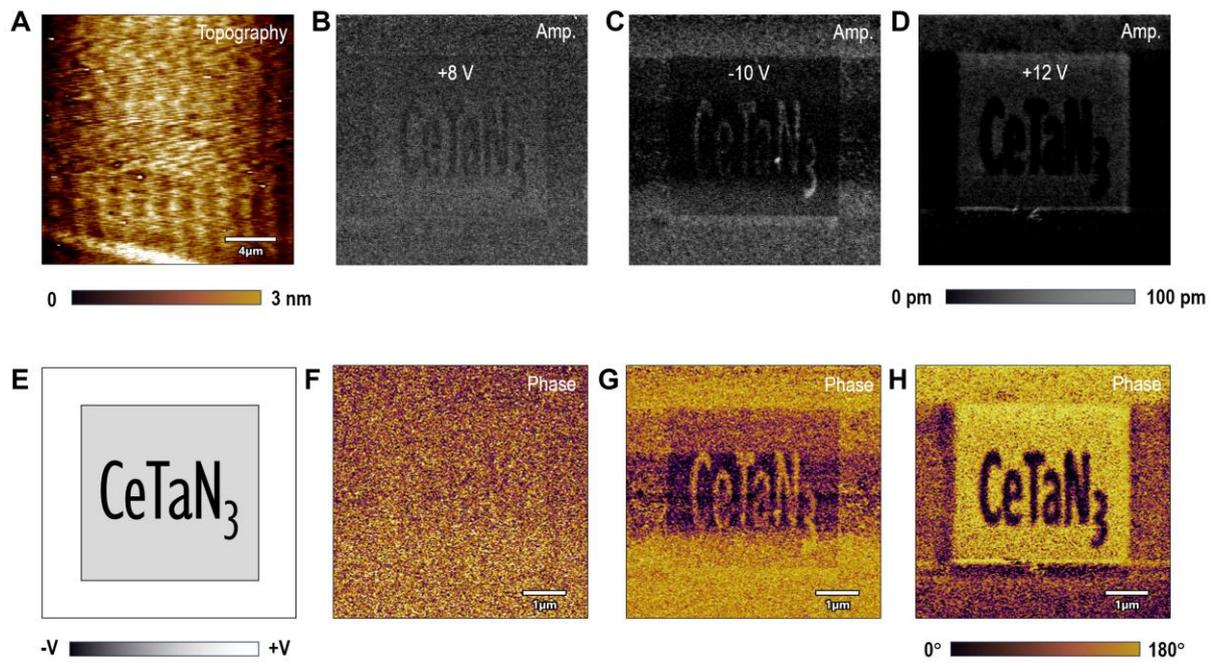

**fig. S19. PFM results on a 15-nm-thick CeTaN$_3$ thin film grown on SrTiO$_3$/Si.** (**A**) Topography of sample, yielding a r.m.s of ~ 2.7 nm. The "CeTaN$_3$" letters were written with opposite tip voltages comparing to the switching voltages poling the square area. (**B** to **D**) PFM amplitude and (**F** to **H**) PFM phase images with gradually increasing tip voltages. (**E**) Schematic of electrical switching patterns.

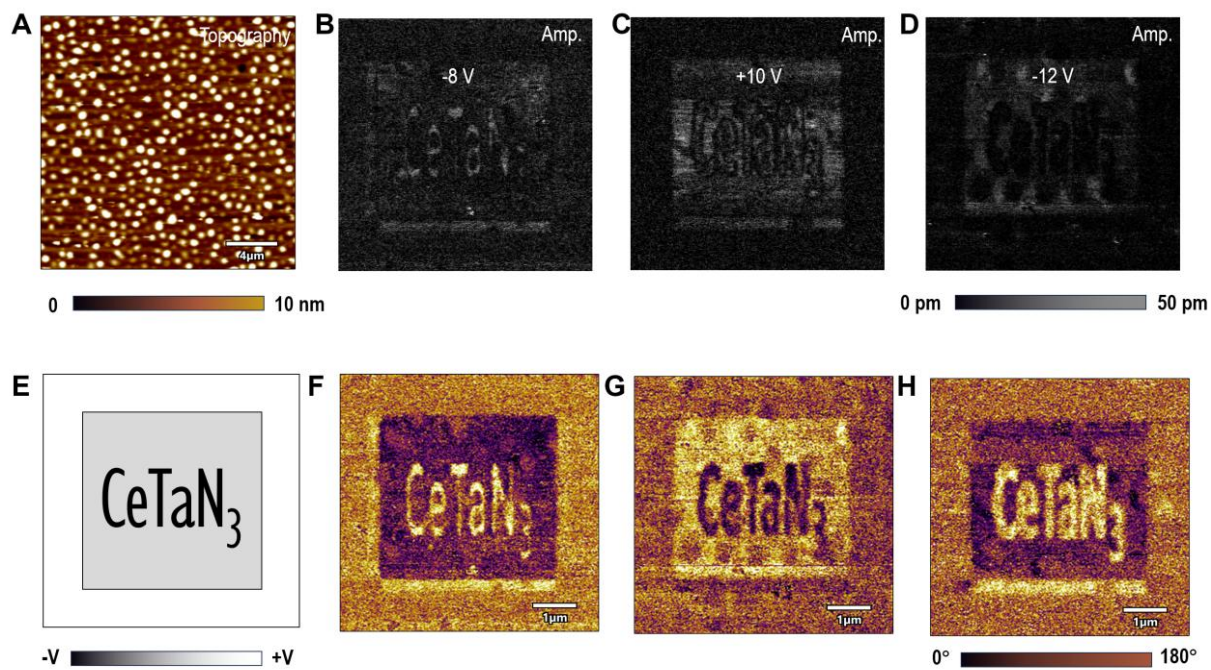

**fig. S20. PFM results on an 80-nm-thick CeTaN$_3$ thin film grown on SrTiO$_3$/Si.** (**A**) Topography of sample, yielding a r.m.s of ~ 8.3 nm. The "CeTaN$_3$" characters were written with opposite tip voltages comparing to the switching voltages poling the square area. (**B** to **D**) PFM amplitude and (**F** to **H**) PFM phase images with gradually increasing tip voltages. (**E**) Schematic of electrical switching patterns.

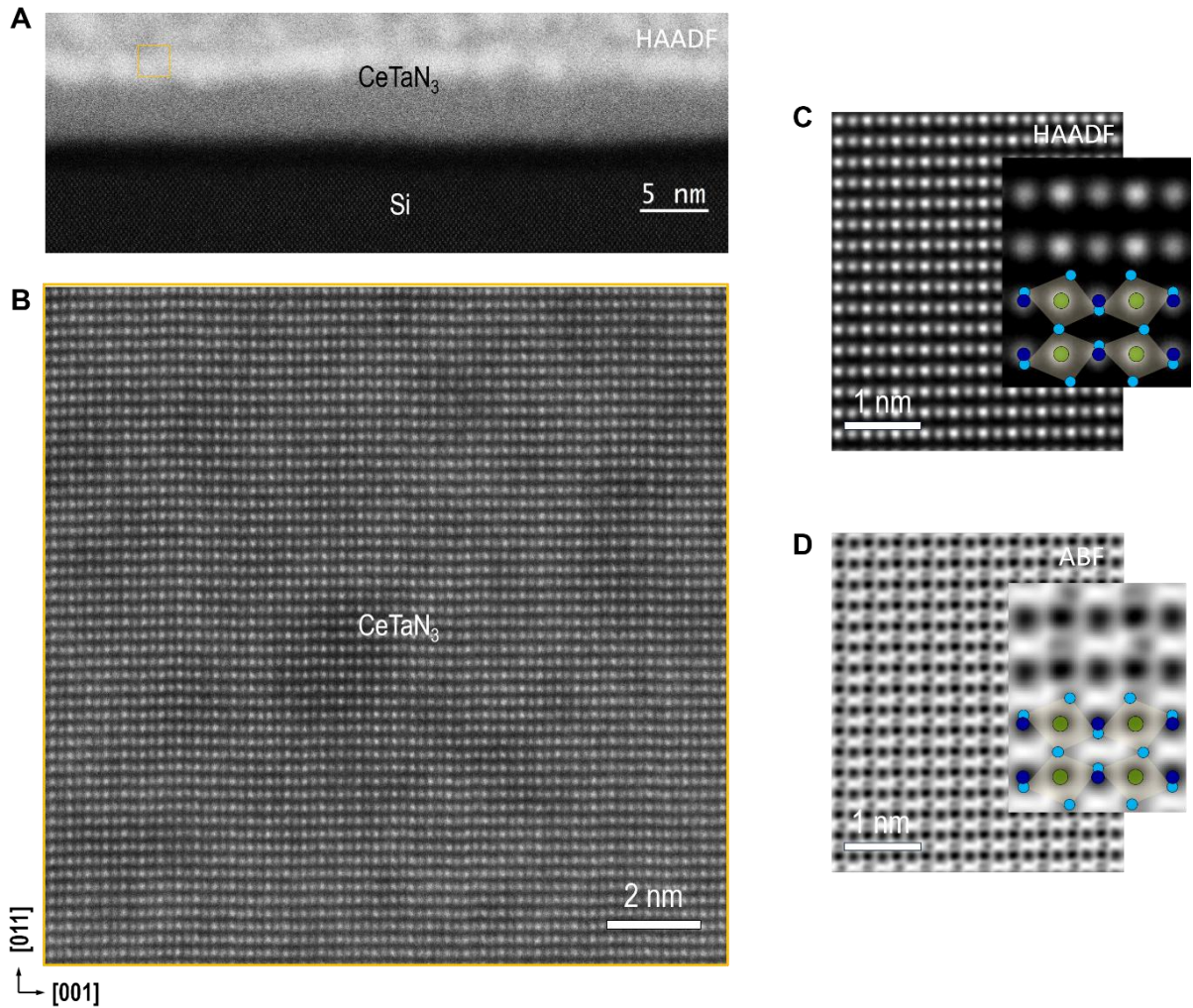

**fig. S21. STEM analysis of CeTaN$_3$ thin films directly on Si.** (**A**) Low magnification cross-sectional STEM-HAADF image at CeTaN$_3$/Si interfaces. Still, we cannot avoid the oxidization of Si at the interfaces at our growth conditions. The thickness of SiO$_2$ layer is approximately 1 nm. (**B**) High magnification STEM-HAADF image of CeTaN$_3$ thin films. The out-of-plane direction is along the [011] orientation. (**C**) STEM-HAADF and (**D**) STEM-ABF images of CeTaN$_3$, from which we could identify the tilted nitrogen octahedral.

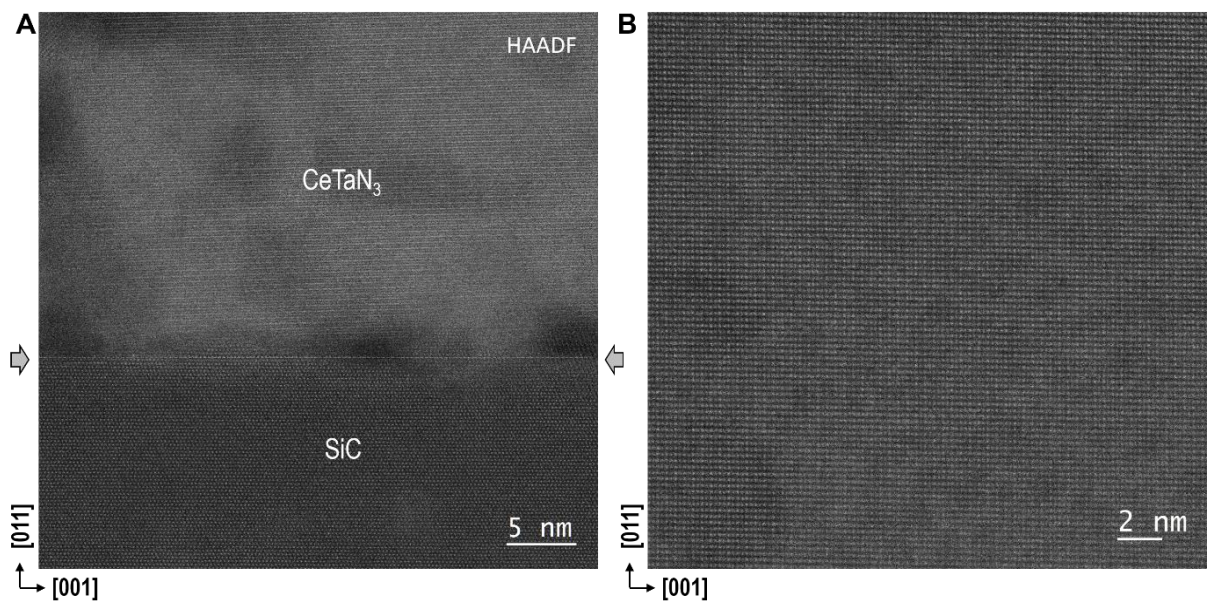

**fig. S22. STEM analysis of CeTaN₃ thin films directly on SiC.** (**A**) STEM-HAADF image at the CeTaN₃/SiC interface region. Arrows and dashed line indicate the position of interface. (**B**) High magnification STEM-HAADF image of CeTaN₃ thin films.

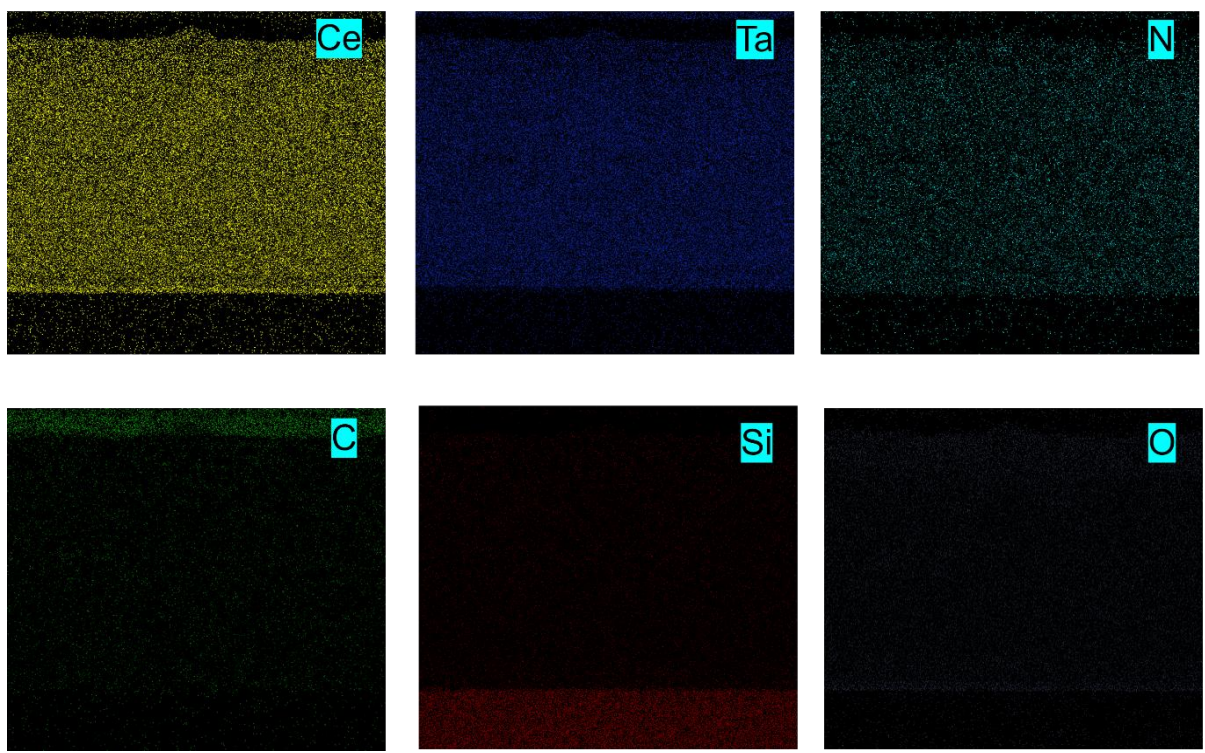

**fig. S23**. **Chemical distribution in CeTaN₃ thin films.** (**A** to **F**) EDX images of Ce, Ta, N, C, Si, and O elements. These results indicate that the CeTaN₃ films exhibit a high-crystallinity and chemical homogeneity without apparent contamination of oxygen.

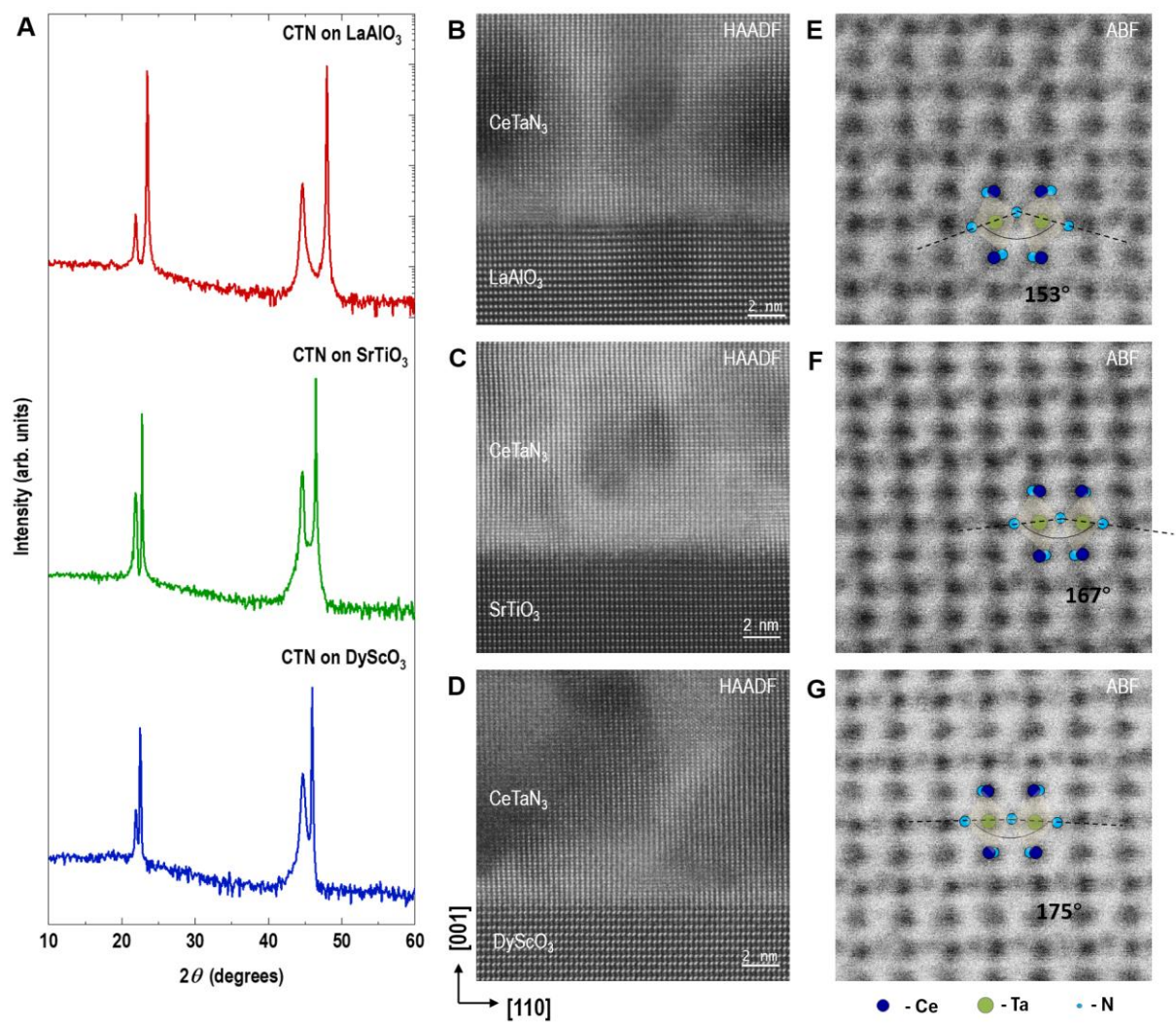

fig. S24. **Structural characterization of CeTaN₃ thin films on different oxide substrates.** (A) XRD θ-2θ scans of CeTaN$_3$ thin films grown on LaAlO$_3$, SrTiO$_3$, and DyScO$_3$ substrates, confirming their epitaxial growth. (**B–D**) HAADF-STEM images of CeTaN$_3$ thin films on LaAlO$_3$, SrTiO$_3$, and DyScO$_3$ substrates, respectively, revealing atomically sharp interfaces. (**E–G**) High-magnification ABF-STEM images of CeTaN$_3$ thin films on LaAlO$_3$, SrTiO$_3$, and DyScO$_3$ substrates, respectively. Insets in each ABF image illustrate the schematic of octahedral tilting, with the corresponding tilt angles indicated.

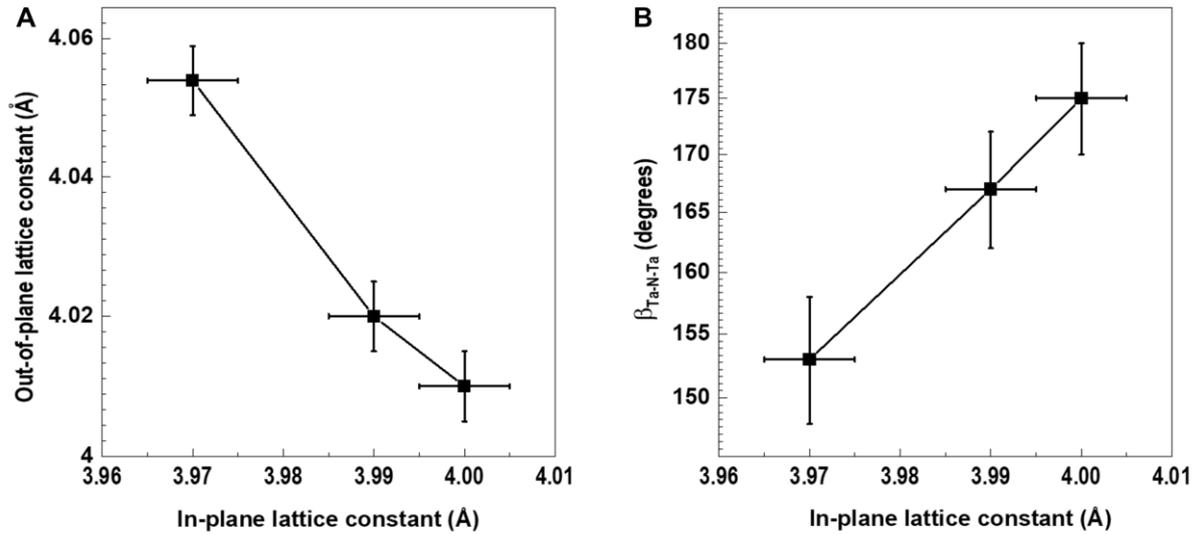

**fig. S25**. **Structural parameter evolution with in-plane lattice constant.** We determined the out-of-plane and in-plane lattice parameters of $CeTaN_3$ thin films using XRD measurements, including θ-2θ scans and reciprocal space mappings (RSMs). (**A**) Out-of-plane lattice parameters and (**B**) $β_{Ta–N–Ta}$ angle as a function of in-plane lattice constants for three $CeTaN_3$ thin films. The results clearly show that the $β_{Ta–N–Ta}$ angle progressively decreases with increasing compressive strain, following the trend from $DyScO_3$ to $SrTiO_3$ and $LaAlO_3$ substrates.

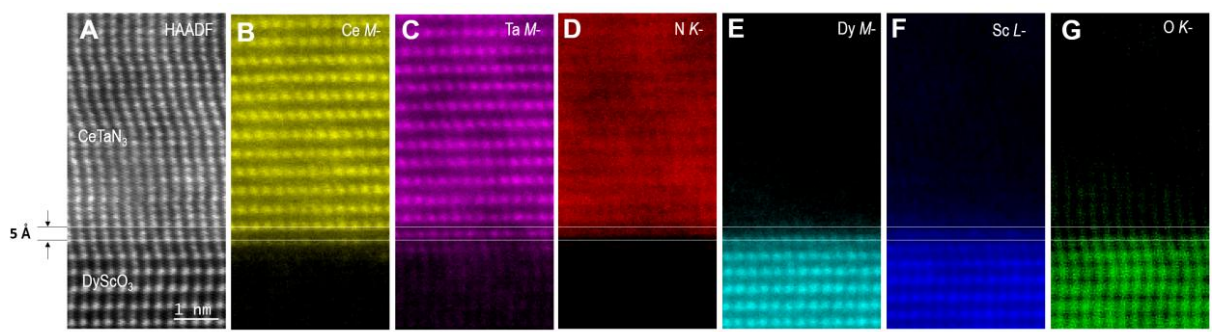

**fig. S26**. **EELS mapping of CeTaN₃ thin films on DyScO₃ substrates.** (**A**) HAADF-STEM image, where the tilt is attributed to thermal drift during STEM imaging. (**B–G**) Atomically resolved EELS maps corresponding to the Ce *M*-, Ta *M*-, N *K*-, Dy *M*-, Sc *L*-, and O *K*-edges. These results confirm the exceptionally high chemical homogeneity of CeTaN₃ thin films, with minimal chemical intermixing at the heterointerface, limited to a length scale of one unit cell. The interface termination is identified as TaN₂–DyO.

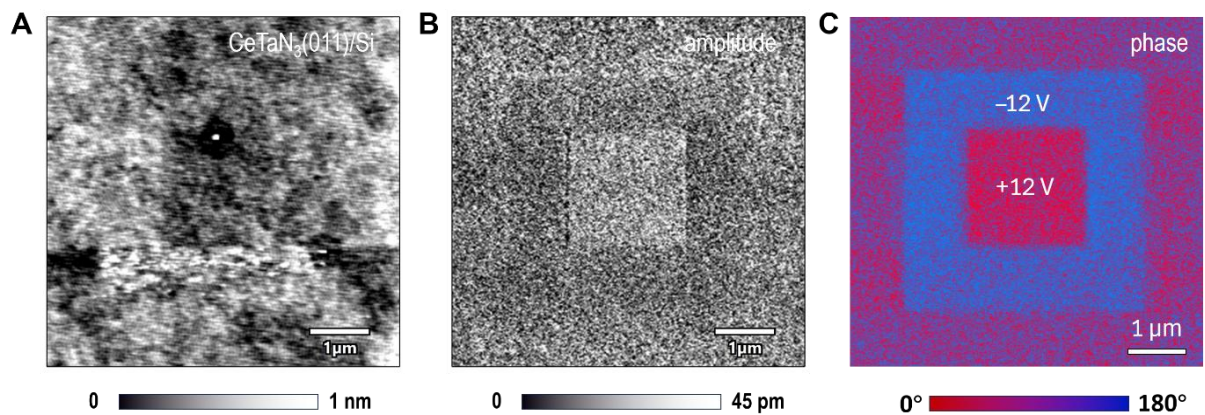

**fig. S27. PFM results of CeTaN₃ thin films directly grown on Si.** (**A**) Topography of CeTaN$_3$/Si. The r.m.s of sample is ~ 1 nm. (**B**) PFM amplitude and (**C**) phase images of CeTaN$_3$/Si. The center and border square area was upward and downward polarized by tip voltages, respectively. the square shapes show reversal phase and amplitude of piezoelectric responses.

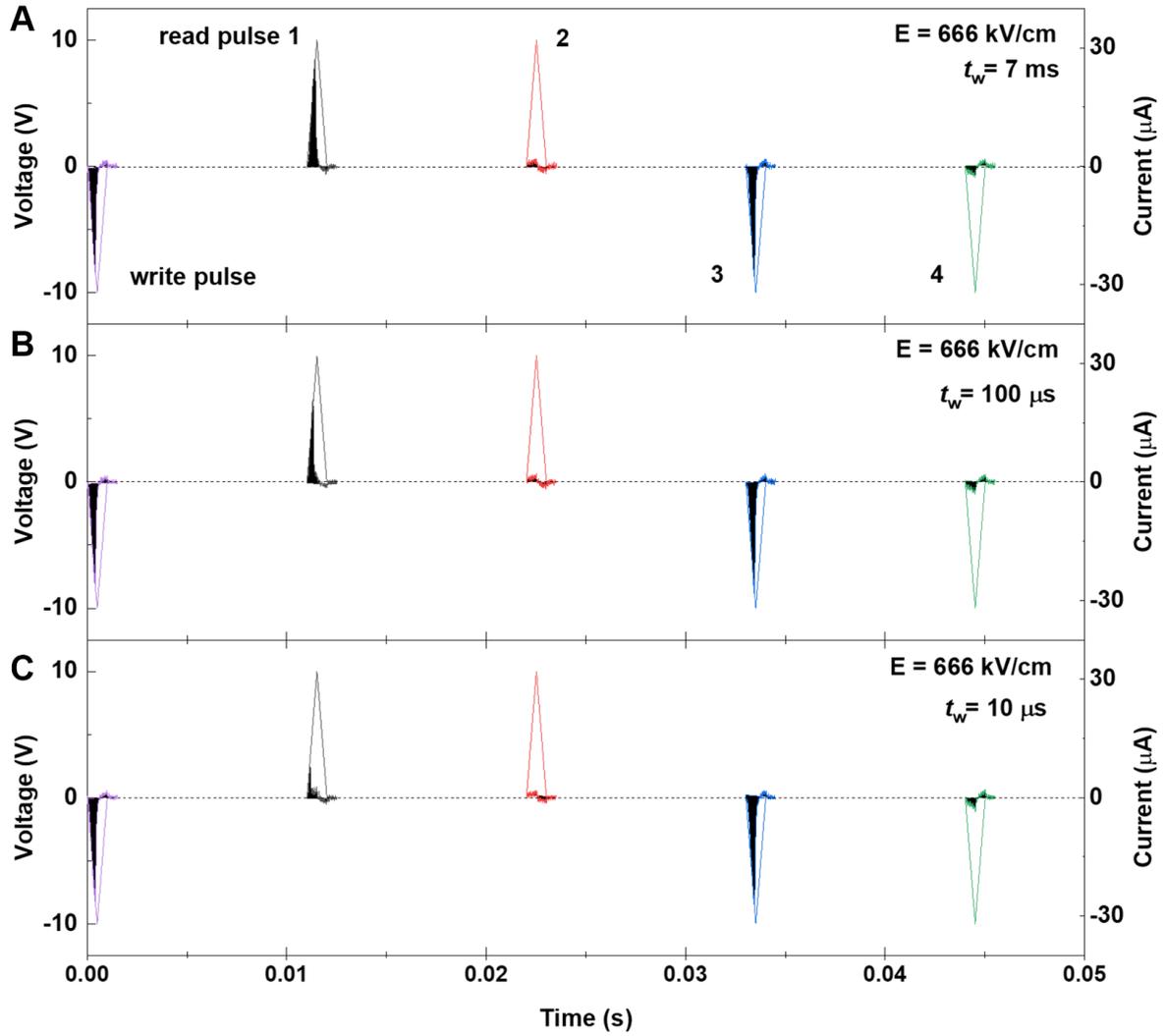

**fig. S28**. **PUND measurements on a Pt/CeTaN$_3$/*p*-Si capacitor.** (A–C) PUND measurements were conducted with a fixed poling electric field (V$_w$ = 10 V, corresponding to 667 kV/cm) at different writing times (*t*$_w$ = 7 ms, 100 μs, and 10 μs), respectively. A sharp difference is observed in the read pulse 1, revealing distinct switching currents.

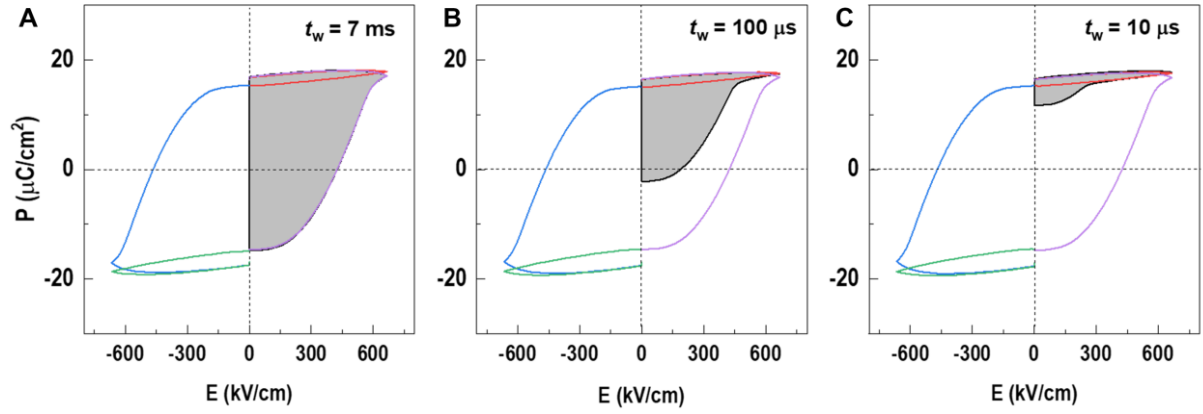

**fig. S29**. **Ferroelectric hysteresis loops of Pt/CeTaN$_3$/*p*-Si capacitor under different writing times (*t*$_w$).** These results are calculated from Fig. S27 A-C. The write and read curves correspond to the curves described in Fig. S27. The shadow areas represent the switched ferroelectric polarizations under different *t*$_w$.

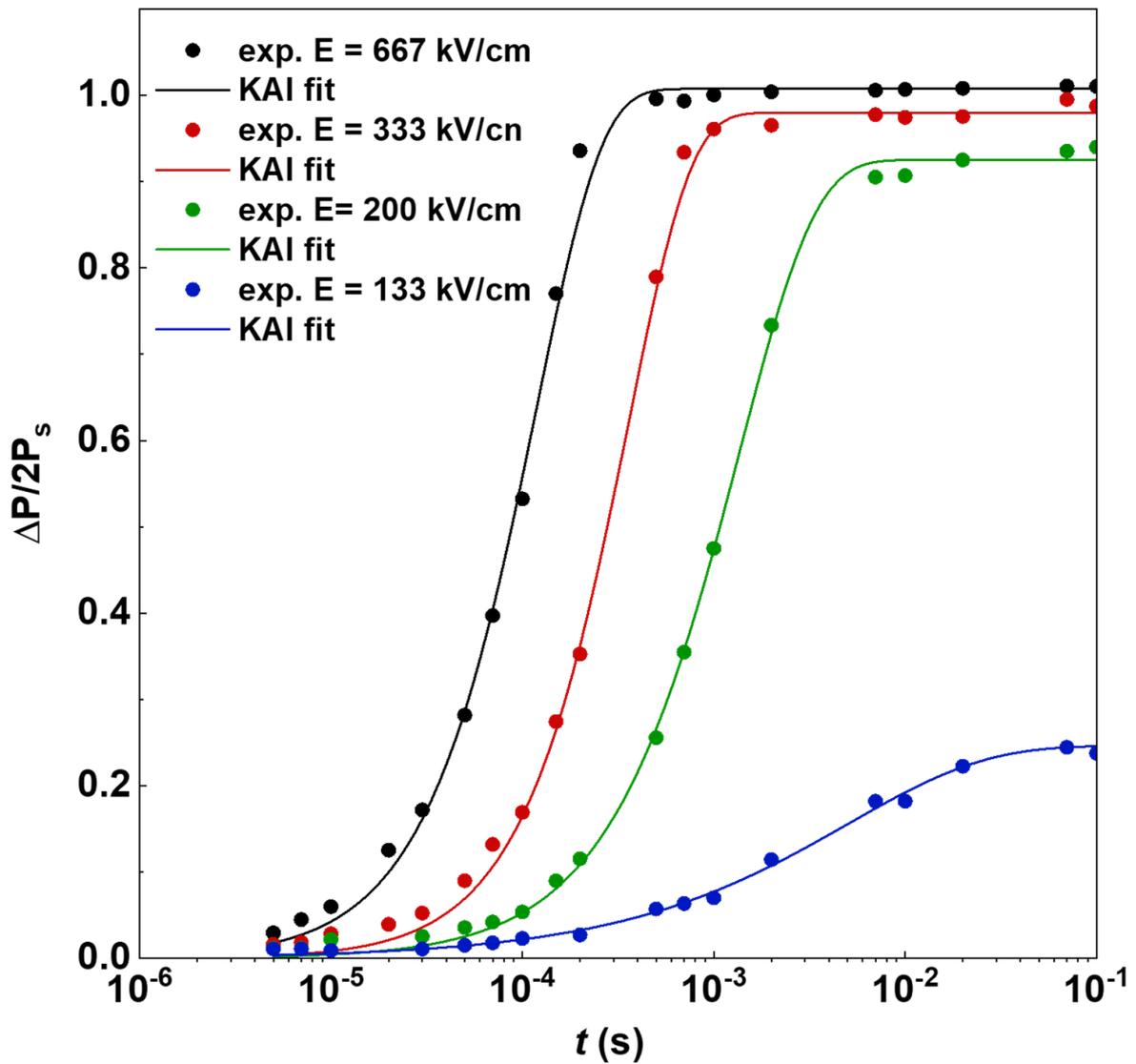

**fig. S30. Ferroelectric switching tests on a Pt/CeTaN$_3$/p-Si capacitor.** The switched polarization (ΔP) as a function of $t$ for four different electric fields. ΔP were obtained by varying the pulse duration ($t$) while keeping the electric field constant. According to the Kolmogorov-Avrami-Ishibashi (KAI) model (*58, 59*), the switched polarization follows the relation ΔP = 2P$_s${1 − exp [−$t$/$t_{sw}$]} for a continuous one-dimensional medium. Fitting the data using the KAI model reveals that the switching time ranges from 1.18 × 10$^{-4}$ s to 0.005 s, depending on the poling fields. Higher electric fields result in faster ferroelectric domain switching.